# Quantifying the Impact of Starspot-crossing Events on Retrieved Parameters from Transit Lightcurves

Catriona A. Murray and Zachory Berta-Thompson
University of Colorado, Boulder, USA
*Received 2025 October 20; revised 2026 March 24; accepted 2026 April 1; published 2026 May 11*

## Abstract

Starspot-crossing events (SCEs) in exoplanet transit lightcurves are becoming increasingly common as we focus on cooler host stars and observe higher precision photometric and spectroscopic lightcurves. In this work, we explore how these events affect our retrievals of transit depths and the accuracy with which we can derive spot properties. We inject and recover synthetic SCEs in photometric lightcurves using `starry`. We find that for high signal-to-noise ratio SCEs, we constrain the spot longitudes tightly (>80% within 1° of the true value) but degeneracies complicate retrieving spot contrasts, radii, and latitudes (within 17%, 19%, and 9° respectively). On average, the difference between injected and recovered transit depths is 0.78% or 78.3 ppm. In most (80%) injections, we recover the transit depth to within 0.6%. For transit depths inflated >1.3% by the transit light source effect (TLSE), fitting for a spot crossing improves the transit depth retrieval over masking the SCE in >95% of cases. However, we find that for spots with small contrasts (<5%) and/or covering fractions (<2%), we are likely to overcorrect for the TLSE, recovering a worse transit depth than simply masking. In addition, even when fitted, we find SCEs can inflate the uncertainties on recovered transit depths significantly, especially for JWST-like precisions. Finally, we present a new method, using SCE observables, of carving out the degenerate spot parameter space that provides informative priors for Markov Chain Monte Carlo sampling, demonstrating this technique on a real SCE observed in Kepler-51d's lightcurve.

## 1. Introduction

In both exoplanet photometry and transmission spectroscopy, we often assume that the host stars' surfaces are homogeneous and temporally constant. However, stars are speckled with evolving surface heterogeneities, such as dark (e.g., spots, pores, or dark faculae on M-dwarfs; B. V. Rackham et al. 2023) and bright (e.g., faculae or plages) magnetic features. When a planet transits in front of its star, these features introduce wavelength-dependent contamination through both unocculted and occulted active regions. This effect is especially problematic for low-mass M-dwarfs, which tend to have enhanced activities and higher spot-covering fractions than earlier spectral types (B. V. Rackham et al. 2018; G. Somers et al. 2020; T. Barclay et al. 2023). Since these cool stars are host to some of our best candidates for atmospheric characterization with state-of-the-art telescopes, such as JWST (J. P. Gardner et al. 2006), this contamination poses a major barrier to accurate planetary inferences—including both bulk parameters (e.g., planet radius) and atmospheric abundances.

The "transit light source effect" (TLSE; B. Rackham et al. 2017; B. V. Rackham et al. 2018) occurs when there are spectral differences between the transit chord and averaged stellar disk, even if the transit cord crosses no active features, due to unocculted features, temperature gradients, or other surface heterogeneities. This effect can inflate (for dark regions) or shrink (for bright regions) our observed transit depths (S. Czesla et al. 2009; Z. K. Berta et al. 2011; D. K. Sing et al. 2011). Spots can also imprint spectral signatures from water and other molecules (H. R. A. Jones et al. 1995) into transmission spectra, entangled with the planet's atmosphere (B. V. Rackham et al. 2018, 2023; A. R. Iyer & M. R. Line 2020). On M-dwarfs faculae may be bright or dark—which can be difficult to disentangle from cool spots—though 3D hydrodynamical simulations predict that very few bright features form on cool stars, with the change in radiative flux largely dominated by dark magnetic features (B. Beeck et al. 2015; B. V. Rackham et al. 2023).

Therefore, for the remainder of this paper we will only consider cool stellar spots, though this work could be applied to hot spots with small adjustments. Several recent JWST observations of rocky exoplanets (e.g., T. Barclay et al. 2023; E. M. May et al. 2023; T. Mikal-Evans et al. 2023; S. E. Moran et al. 2023) were unable to differentiate between absorption from the planet's atmosphere and from stellar spots. To correct for the TLSE, we often model active regions and quiescent photosphere with stellar spectral components of different temperatures, but M-dwarf stellar models (e.g., T. O. Husser et al. 2013; A. R. Iyer et al. 2023) are discrepant (O. Lim et al. 2023; B. V. Rackham & J. de Wit 2024) and poorly describe empirical spectra (H. R. Wakeford et al. 2019; L. J. Garcia et al. 2022). To robustly interpret exoplanet atmospheres, we need direct observations of active regions on exoplanet host stars.

Occulted spots cause chromatic flux bumps directly in the lightcurve (e.g., A. V. R. Silva 2003; F. Pont et al. 2007; M. Rabus et al. 2009; R. Sanchis-Ojeda & J. N. Winn 2011; D. K. Sing et al. 2011). Joint modeling of these features alongside planetary transits (e.g., K. F. Huber et al. 2010; D. K. Sing et al. 2011; F. Pont et al. 2013; L. Mancini et al. 2017; G. Scandariato et al. 2017; N. Espinoza et al. 2019; M. Fournier-Tondreau et al. 2024) can provide insights into the sizes and temperatures of spots, as well as the covariances between spot and planet parameters (B. V. Rackham et al. 2023). A number of spot- and planet-transit-fitting tools exist for this purpose, including `starry` (R. Luger et al. 2019), FLECK (B. Morris 2020), and `spotrod` (B. Béky et al. 2014).

However, these starspot-crossing events (SCEs) can complicate the retrieval of transit parameters, including transit depth, limb-darkening (introducing slopes in visible and infrared transmission spectra; X. Alexoudi et al. 2020), and other planetary parameters (S. C. C. Barros et al. 2013). SCEs have already been seen in a number of recent JWST observations (G. Fu et al. 2022; M. Fournier-Tondreau et al. 2024; J. E. Libby-Roberts et al. 2026) and, although spot crossings are common on M-dwarfs, there is little consensus on the best treatment. Often, they are simply masked (e.g., F. Pont et al. 2008; N. Narita et al. 2013; G. Fu et al. 2022); however, this can leave artifacts in the lightcurve and the transit depth inflation due to the TLSE must still be accounted for afterwards. SCEs are often unpredictable due to largely uncharted and evolving stellar surfaces. Currently, 3D magnetohydrodynamic (MHD) stellar models for M-dwarfs are difficult to verify because we cannot easily resolve extrasolar surfaces, which limits our ability to model the structure and behavior of these features. Therefore, it is challenging to predict how SCEs will affect our planet retrievals, particularly when considering how to conservatively account for such events when proposing for telescope time.

Additionally, while transit scans over active regions can, in principle, constrain spot sizes, locations, and temperatures, there are several well-known, though poorly characterized, degeneracies to consider. For example, the photometric signature of a small spot located near the center of the transit chord can resemble that of a larger spot positioned farther from the chord—demonstrating the strong degeneracy between spot radius and latitude. Similarly, a cooler smaller spot can result in a similar SCE as a larger warmer spot; we have further degeneracy with spot temperature. When considering occulted spots, we also have to contend with whether the spot is projected "above" or "below" the transit chord. These degeneracies, which affect both occulted and unocculted spots, can be difficult to resolve with single transits and photometry alone, though repeated transits of the same planet over its star's rotation period can help constrain the sizes and locations of spots (e.g., HAT-P-18b; B. M. Morris et al. 2017).

Existing spot-modeling tools attempt to fit degenerate spot parameters (e.g., size, location, and temperature) directly by sampling the full, highly multimodel parameter space. In such landscapes, gradient-based samplers and optimizers often become trapped in local likelihood maxima, fail to explore disconnected regions of parameter space, and require substantial computational effort (J. Dunkley et al. 2005). However, the observable characteristics of the SCE in the lightcurve (time, duration, size), when combined with simple geometric considerations, can already exclude large families of spot solutions. While spot degeneracies are widely recognized, it remains difficult to quantify the full range of physically consistent spot properties when inference relies solely on posterior samples that may explore only isolated regions of parameter space. In this work, we present a new practical framework that uses lightcurve observables to map and prune the degenerate space prior to fitting the SCE. Rather than relying exclusively on sampling to reveal degeneracy structure, we use the geometry of the SCE itself to better understand the degeneracy space and to provide physically motivated priors for sampling.

While multiwavelength spot-crossing observations can provide useful leverage to constrain spot temperature and disentangle it from geometry, these data are not always available and stellar models for M-dwarfs remain unreliable. In practice, most spectroscopic spot-crossing fits involve first fitting the white-light transit to fix the geometric spot parameters and then holding these fixed when fitting spectroscopic lightcurves to extract contrast spectra. The accuracy of those multiwavelength corrections therefore hinges on how well the spot's position and size can be determined from a single wavelength. Robustly characterizing the spot geometry then improves the fidelity of the contrasts inferred in spectroscopic channels. Once we have derived the contrast spectrum from occulted spots, we can then easily account for the unocculted spots (assuming they share a common temperature) and correct for the contamination in the transmission spectra directly (only needing to fit for the unocculted spot-covering fraction) without relying on imperfect stellar atmosphere models.

Therefore, in this paper we use white lightcurves to explore the key questions of:

1. If there is an SCE in my lightcurve, should I mask or model it? And, how will that decision affect the recovered transit depths and uncertainties?
2. How well can spot parameters be constrained from a single event and wavelength?
3. Which degenerate spot scenarios are consistent with an SCE?

In this work we use the `chromatic_fitting` (with `starry`) tool, described in Section 2, to simultaneously model planet transits and SCEs. We inject simulated SCEs (Section 3) into transit lightcurves and recover the transit depths and spot parameters (Section 4). For a handful of SCEs, we perform Markov Chain Monte Carlo (MCMC) sampling to extract the posterior distributions in Section 5. We conclude with an exploration of the spot-crossing degeneracy problem: presenting our new framework for pruning unphysical parameter space in Section 6 and applying to a real SCE in the JWST transit of Kepler-51d in Section 7.

## 2. Chromatic_fitting

When extracting information from an exoplanet's transit, we cannot consider the planet in isolation. We are observing entangled spectral signatures from the planet's atmosphere, the star's active surface, and uncorrected instrumental effects. Therefore, we adopt a simultaneous fitting approach to disentangle these signals, capture parameter covariances, and yield more robust uncertainties.

We employ the open-source Python tool chromatic_fitting (C. Murray et al. 2025)[1], which can model lightcurve features imprinted by a number of sources—planetary, stellar or instrumental—at the same time. chromatic_fitting utilizes the framework of chromatic (Z. Berta-Thompson et al. 2025),[2] which defines spectroscopic lightcurve objects as Rainbows, with wavelength, time, flux, and uncertainty attributes. chromatic_fitting is designed to be fast, flexible, and user-friendly, enabling the combination of different modular models (including wrappers for commonly used tools) to fit multiwavelength photometric data and extract transmission and emission spectra of planetary atmospheres. Though optimized for JWST (applied in E.-M. Ahrer et al. 2023; JWST Transiting Exoplanet Community Early Release Science Team et al. 2023; J. E. Libby-Roberts et al. 2026; P. Wachiraphan et al. 2025), this tool is broadly applicable to any photometric or spectroscopic planet observations—including from space-based, ground-based, or multiple facilities. This work marks the first comprehensive validation of chromatic_fitting for transits with starSCEs.

### 2.1. Models

There are a number of model modules available within chromatic_fitting. Some models are built on existing tools, such as exoplanet (D. Foreman-Mackey et al. 2021), starry, and celerite2 (D. Foreman-Mackey et al. 2017; D. Foreman-Mackey 2018), and some are coded within chromatic_fitting itself. The planet models available are: TransitModel (exoplanet), EclipseModel (secondary eclipse, starry), PhaseCurveModel (starry), TransitSpotModel (transit with occulted spot, starry), and TrapezoidModel. The systematic models available are: GPModel (Gaussian process, celerite2), PolynomialModel, StepModel, ExponentialModel, and SinusoidModel. Any number of these models can be combined flexibly to model the planet, star, and/or instrumental systematics in one fit.

### 2.2. Flexible and Multiwavelength Fitting

chromatic_fitting was designed to fit multiwavelength lightcurves to fully exploit the wide wavelength coverage of facilities such as JWST. It has the ability to fit wavelengths individually (separate fitting), simultaneously (simultaneous fitting), as well as to fit the white lightcurve. In separate fitting, no parameters are shared across wavelength; however, in simultaneous fitting there is the option to fit shared parameter values across all wavelengths (e.g., semimajor axis) or to fit a different value for each wavelength (e.g., transit depth). While both chromatic_fitting and chromatic are tools designed for multiwavelength analyses, in this paper we will focus on single wavelengths (separate fitting) only, discussed in more detail in Section 3.1.

## 3. A Sample of Starspot-crossing Events

A number of spot occultation tools available to model SCEs. To our knowledge, they all follow one of three conventions for defining stellar spots: (a) spots are hard-edged circles (e.g., fleck; B. Morris 2020, spotrod; B. Béky et al. 2014, STSP; B. M. Morris et al. 2017), (b) a smooth expansion of spherical harmonics (e.g., starry; R. Luger et al. 2019), or (c) pixelated shapes (e.g., ECLIPSE; A. V. R. Silva 2003, SOAP-T; I. Boisse et al. 2012, PRISM; J. Tregloan-Reed et al. 2013, KSint; M. Montalto et al. 2014, ellc; P. F. L. Maxted 2016, TOSC; G. Scandariato et al. 2017, PyTranSpot; I. G. Juvan et al. 2018, pixel mapping in starry, StarSim2; A. Rosich et al. 2020, spotter[3]).

For this work, we elected to use starry, which uses combinations of spherical harmonics to create surface maps with a resolution set by the number of spherical harmonic orders, though our methods could be repeated with any tool. starry also accounts for limb darkening and the subsequent effect of spots near or on the stellar limb. We will discuss where differences arise between the spherical harmonic and hard-edged circle approaches in this (see Section 3.3) and later (see Sections 6.3 and 7.2) sections.

### 3.1. Single-wavelength Lightcurves

In this work, we consider only single-wavelength transits (equivalent to a broadband or white lightcurve) for three main reasons. (1) We aim to determine how well stellar contamination can be constrained without stellar model assumptions. We generate spot contrasts from a uniform distribution, ignoring any underlying spectral character. (2) Generally, we assume that spot size and position are wavelength-independent (B. V. Rackham et al. 2018, 2019; G. Bruno et al. 2021; M. C. Schutte et al. 2023). Therefore, the best possible estimates for these parameters are typically derived from a white lightcurve, before being fixed in spectroscopic fits. Since the contrast spectrum is tied to the spot geometry, accurate single-wavelength parameters are essential for inferring accurate contrasts and correcting for spots in multiwavelength datasets and transmission spectra. (3) In practice, observers may only have photometry available, so it is important to understand what constraints are possible without spectral information. By varying spot contrast and data uncertainty in our sample, we hope that our results can be generalized to any wavelength.

### 3.2. Generating a Sample of Starspot-crossing Events

To create a large sample of single SCEs (one occulted spot in one transit) we first generated a simple star with a planet transiting on a circular orbit using starry. We defined a star with $R_* = 1\,R_\odot$, $M_* = 1\,M_\odot$, a rotation period of 1000 days (to negate stellar spot rotation over the transit duration), inclined at $i_* = 90°$, and quadratic stellar limb-darkening coefficients of [0.02, 0.2]. Its transiting planet has radius $R_{\rm p} = 0.1\,R_\odot$, an orbital period $P = 1$ day, and an impact parameter $b = 0$ (i.e., it transits the center chord of the star). By setting $b = 0$, we can exploit the symmetry of the transit and only consider spot crossings on one hemisphere of the star (see Figure 2).

We simulated a "typical" single transit lightcurve for this system with equal time in and out of transit (the system and orbital parameters produce a ≈2 hr transit duration resulting in ≈4 hr lightcurve) and a cadence chosen to give 100 data points in transit and 100 out of transit, with approximately 10 points during ingress and egress (cadence of ≈73 s). We then

[1] https://github.com/catrionamurray/chromatic_fitting
[2] https://github.com/zkbt/chromatic
[3] https://github.com/lgrcia/spotter

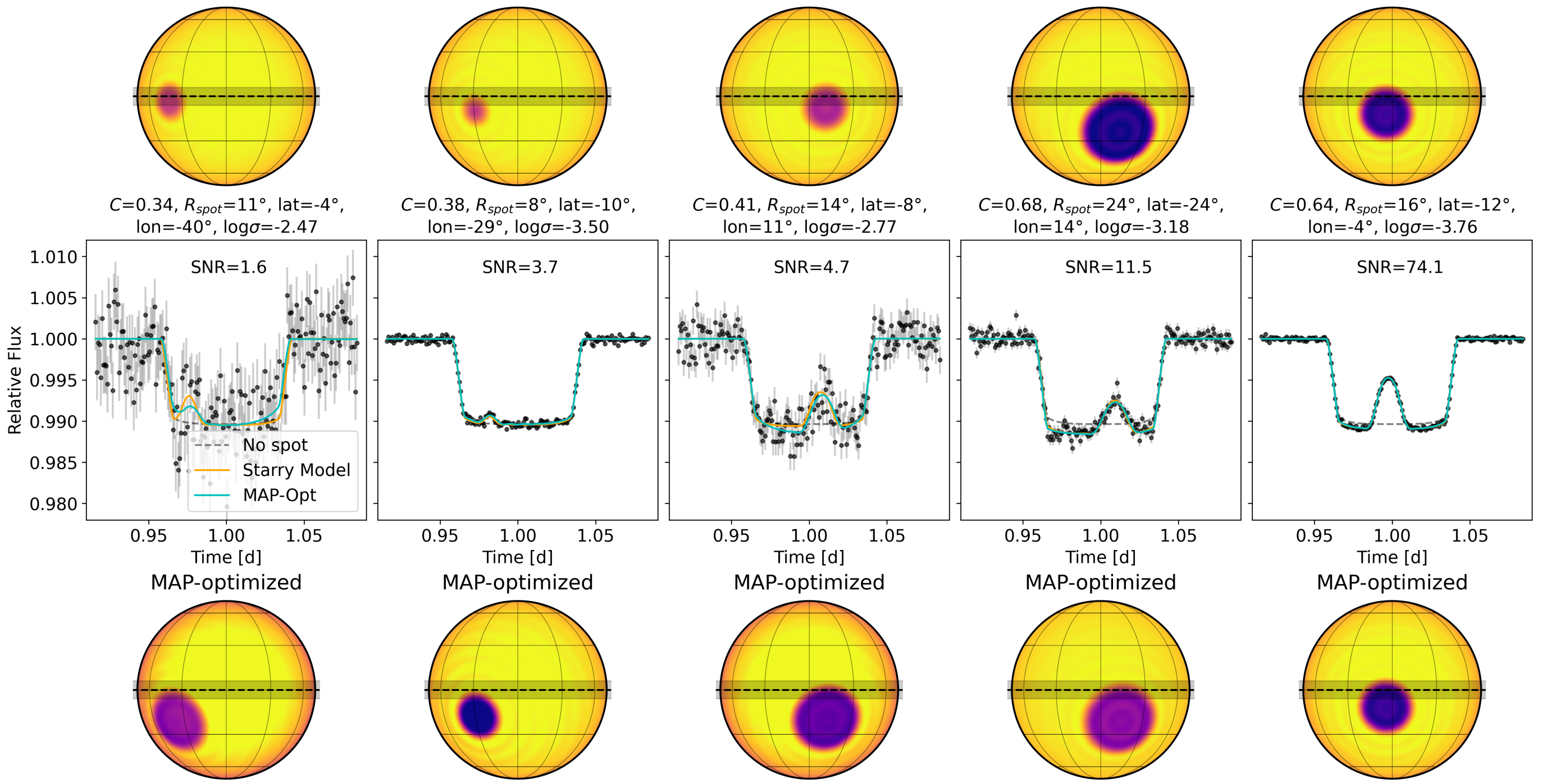

**Figure 1.** We present five samples from Section 3.2 to demonstrate a range of lightcurve uncertainties and spot parameters. Upper: The `starry` stellar surface maps for each injection scenario with quadratic limb-darkening and the spot. There are slight ringing artifacts from the choice to reduce the smoothing parameter, as discussed in Section 4.1. The transit chord is shown by the shaded gray region on each map. Middle: The transit lightcurves in each case, with injected Gaussian uncertainties, where the amplitudes are drawn from the $\log \sigma$ prior in Equation (1). The underlying model generated by `starry` is shown in orange, the same transit without spot crossing or TLSE in dotted gray, and the recovered maximum a posteriori (MAP)-optimized result shown in blue. Lower: The `starry` stellar surface maps for the MAP-optimized solution in each case.

modified this transiting system to include 1000 simulated star SCEs with the following distributions:

$$\begin{aligned} C &\sim U(0.05,\ 1) \\ \log(R_{\rm spot}) &\sim U(5,\ 45)^\circ \\ y &\sim U(-0.8,\ 0)R_* \\ x &\sim U(-\sqrt{R_*^2 - y^2},\ \sqrt{R_*^2 - y^2}) \\ \log(\sigma) &\sim U(-5,\ -2), \end{aligned} \quad (1)$$

where $C$ is the spot contrast (where a spot with $C = 0$ has the same temperature as the photosphere and $C = 1$ is a perfectly dark spot), $R_{\rm spot}$ is the angular radius of the spot, $x$ and $y$ are the $x$- and $y$-position on the visible projected stellar disk (where $x = y = 0$ is the center of the star, the $x$-axis is along the transit chord with the $y$-axis perpendicular), and $\sigma$ is the per-cadence fractional flux uncertainty in the lightcurve. Sampling in $x$ and $y$, rather than latitude ($\phi_{\rm spot}$) and longitude ($\lambda_{\rm spot}$), produces a more uniform distribution of starspots across the projected stellar disk and avoids a disproportionate "build-up" at the stellar limbs (where we expect recovery to be more challenging). $y = -0.8R_*$ is the lowest $y$-position for which a spot of $R_{\rm spot} = 45°$ will overlap with the transit chord. We sampled in $\log(R_{\rm spot})$ to counter that large spots can be occulted at greater latitudes than small spots. As a result our sample is slightly biased toward smaller spots, though these appear to be more common in the literature (e.g., L. Mancini et al. 2013, 2014; J. R. Barnes et al. 2015; B. M. Morris et al. 2017; J. E. Libby-Roberts et al. 2023; F. Biagiotti et al. 2024). The lower radius limit of 5° was chosen because it is $\approx R_{\rm p}$, the smallest resolution with which we can map the stellar surface. It is possible, even likely, that spots with radii $<5°$ exist; however, in this work we focus on spots that are large enough to temporally resolve. We chose the large upper limit of 45° to allow for the possibility of very large spots, however, this is likely unphysical for most systems. Five of the 1000 samples, with a range of parameters, are demonstrated in Figure 1 and the overall sample is shown in Figure 2.

### *3.3. Determining the Spot-crossing Properties*

We perform this section's injection and recoveries using `starry`, which requires the star–planet parameters and spot radius, latitude, longitude, and contrast to be defined. `starry` implements spots as a spherical harmonic expansion of a top-hat function which is subject to ringing artifacts at finite harmonic orders (Gibbs phenomenon). To mitigate this, the spots are Gaussian-smoothed on the stellar surface. While this suppresses ringing, it also can broaden the spots, reducing the amplitude and altering the shape of their lightcurve signatures—both of which are directly tied to the spot properties. Furthermore, spots off the transit chord can still imprint structure onto the lightcurve. As a result, determining useful quantities (e.g., spot-crossing times/cadences, spot-coverage fraction, TLSE depth inflation) is not straightforward. Therefore, we also define a simple geometric spot-crossing model (Section 3.3.1), which we use to extract spot coverage (Section 3.3.2), TLSE depth inflation (Section 3.3.1), and signal-to-noise ratio (SNR) for each SCE (Section 3.3.3).

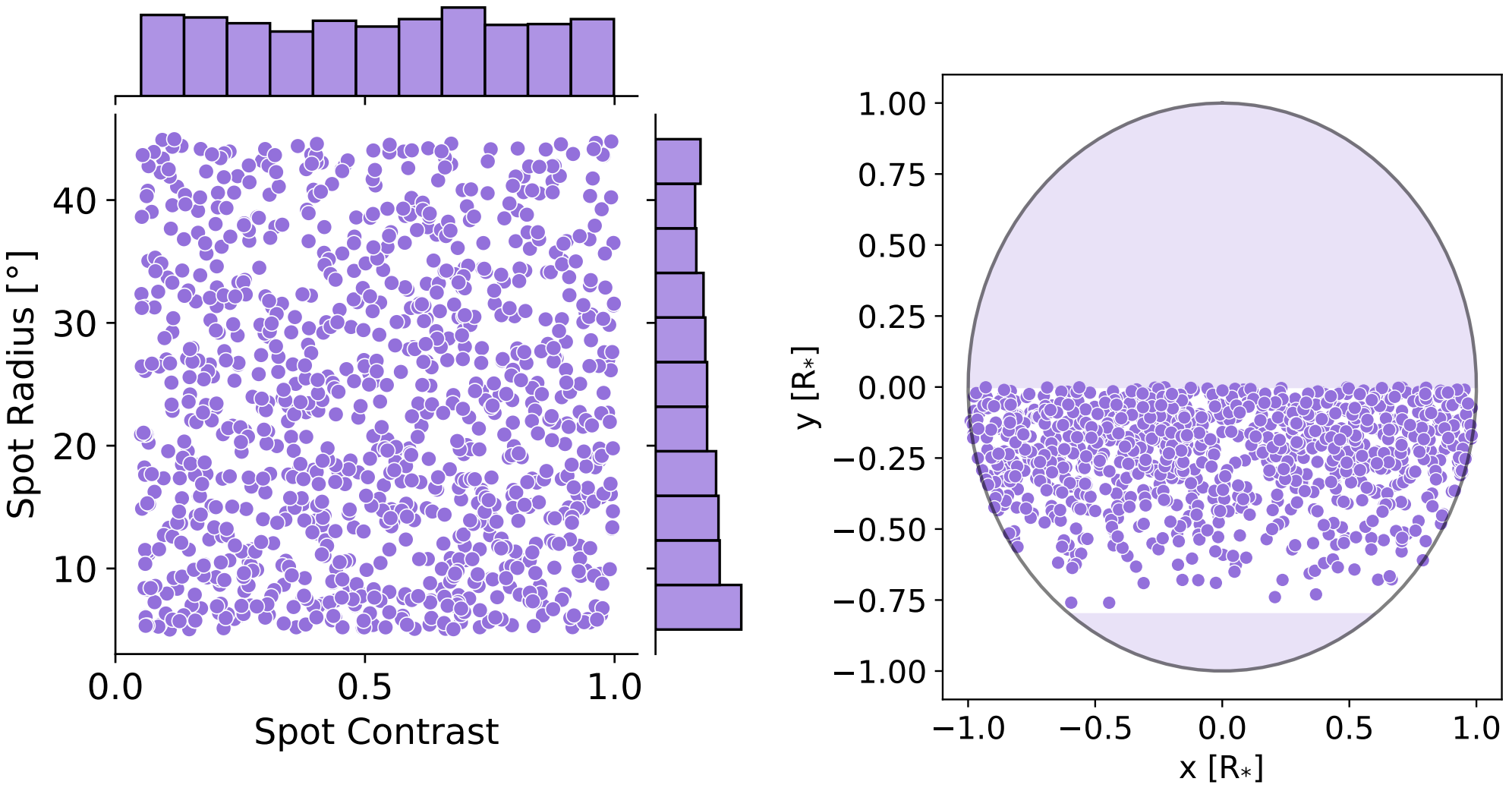

**Figure 2.** The 1000 injected spot samples. Left: Radius of the spot (in degrees), $R_{\rm spot}$, against the contrast between spot and quiescent photosphere. Right: The projected location of the spot centers on the surface of the star. The top half of the star is shaded because we only inject spots into one half to avoid the symmetrical degeneracy in recovery. Below $y \leqslant -0.8$ is also shaded because the largest spot we inject, $R_{\rm spot} = 45°$, would not cross the transit chord.

*3.3.1. Generating a Spot-crossing Model*

Following the formalism in B. Rackham et al. (2017) and B. V. Rackham et al. (2018), we calculate the instantaneous transit depth $D(t)$ during the SCE as

$$D(t) = 1 - \frac{F_{\rm in}(t)}{F_{\rm out}}, \tag{2}$$

where $F_{\rm in}(t)$ is the flux during transit at time $t$ and $F_{\rm out}$ is the (assumed constant) flux outside of transit. We can assume that the total wavelength-dependent ($\lambda$) spectrum of the star $S$ is comprised of the spectral contributions of the unspotted photosphere, $S_{\rm phot}$, and spots with covering fraction $f$, $S_{\rm het}$, to give

$$S(\lambda) = fS_{\rm het}(\lambda) + (1 - f)S_{\rm phot}(\lambda). \tag{3}$$

We assume here this is temporally constant across our single observation, though in reality $S$ may vary during a transit duration if the star is rotating rapidly.

We can then write an equation for the spectrum of the star being eclipsed by the planet at time $t$, $S_{\rm ecl}(t)$

$$S_{\rm ecl}(t, \lambda) = g(t)S_{\rm het}(\lambda) + (1 - g(t))S_{\rm phot}(\lambda), \tag{4}$$

where $g(t)$ is the time-dependent fraction of the planet's shadow that is covered by spots. If $g(t) = f$ (what is behind the planet is the same as the average visible stellar disk), then $S_{\rm ecl}(t, \lambda)$ reduces to $S(\lambda)$. Expressing the fluxes in terms of these spectral components

$$F_{\rm out}(\lambda) = S(\lambda)\pi R_*^2, \tag{5}$$

$$F_{\rm in}(\lambda) = S(\lambda)\pi R_*^2 - S_{\rm ecl}(t, \lambda)\pi R_{\rm p}^2, \tag{6}$$

assuming the planet is totally blocking the star behind its shadow. Therefore, we can rearrange Equation (2) to

$$D(t, \lambda) = \frac{1 - g(t)\left[1 - \frac{S_{\rm het}(\lambda)}{S_{\rm phot}(\lambda)}\right]}{1 - f\left[1 - \frac{S_{\rm het}(\lambda)}{S_{\rm phot}(\lambda)}\right]} D_{\rm true}, \tag{7}$$

$$= \frac{1 - g(t)C(\lambda)}{1 - fC(\lambda)} D_{\rm true}, \tag{8}$$

where $D_{\rm true}$ is the true transit depth, $\left(\frac{R_{\rm p}}{R_*}\right)^2$ and $C(\lambda)$ is the wavelength-dependent spot contrast. The contamination in this case is equivalent to the contamination spectrum, $\epsilon(\lambda) = \frac{1}{1 - fC(\lambda)}$, for unocculted features defined in B. V. Rackham et al. (2018) multiplied by an extra time-dependent factor

$$D(t, \lambda) = [1 - g(t)C(\lambda)]\, \epsilon(\lambda) D_{\rm true}(\lambda). \tag{9}$$

Therefore, as $D(t, \lambda) = \epsilon(\lambda)D_{\rm true}(\lambda) + \Delta D(t, \lambda)$, the change in depth due to the spot crossing is

$$\Delta D(t, \lambda) = -g(t)C(\lambda)\epsilon(\lambda)D_{\rm true}(\lambda). \tag{10}$$

We consider a few quick sense checks:

1. If the planet crosses a region of the star which has the same spot-covering fraction as the entire stellar surface at time $t_i$: $g(t_i) = f$, there is no TLSE and $D(t)$ simplifies to $D_{\rm true}$.
2. If the planet does not occult an active feature at time $t_i$: $g(t_i) = 0$ and Equation (9) reduces to the transit depth contaminated by unocculted spots, $\epsilon D_{\rm true}$.
3. If the planet's shadow falls entirely within the spot: $g(t_i) = 1$ and the transit depth depends on the contrast $C_\lambda$. If the active features are perfectly dark ($S_{\rm het}(\lambda) = 0$, $C_\lambda = 1$), the transit depth falls to 0 and the flux returns to the out-of-transit baseline.
4. If the planet's shadow covers a region with more spots than the average stellar surface: $g(t_i) > f$ and the transit depth decreases, $D(t) < D_{\rm true}$, since it appears at that time as if the planet is eclipsing a "cooler" star.

We note that in this spot-crossing formalization we neglect second-order variations in limb-darkening within spots; assuming limb-darkening affects the star, spot, and transit chord in approximately the same way.

#### 3.3.2. Calculating the Spot-coverage Fraction

To calculate the flux impact of the spot crossing, we need the projected spot-covering fraction, $f$. We assume that spot boundaries can be modeled as circles on the surface of a sphere, calculating their ellipsoidal projections. We made use of an existing Python tool, SHAPELY (S. Gillies et al. 2024), which takes in a list of coordinates to create closed loop objects called *LinearRing*s which can be turned into *Polygon*s with a very high number of sides. Once we have created a *Polygon*, we can use SHAPELY to calculate its area and the intersected area with other *Polygon* objects. We model the planet, spot, and star as *Polygon*s to calculate the ratio between the projected areas of the spot and star (i.e., spot-covering fraction $f$) and spot and instantaneous overlap between the planet's shadow and projected spot, $g(t)$ in Equation (9), at every time point.

#### 3.3.3. Signal-to-noise Ratio of Starspot-crossing Events

Similarly to how we calculated the SNR of a planetary transit, we can calculate the SNR of each spot occultation as follows:

$$\mathrm{SNR} = \frac{\Delta D_{\mathrm{spot}}}{\sigma_{\mathrm{spot}}}, \quad (11)$$

where $\Delta D_{\mathrm{spot}}$ is the height of the spot-crossing shape in the lightcurve and $\sigma_{\mathrm{spot}}$ is the uncertainty across the spot crossing, defined as

$$\sigma_{\mathrm{spot}} = \sqrt{\sigma_i^2 + \sigma_o^2}, \quad (12)$$

$$= \sigma\sqrt{\frac{1}{N_i} + \frac{1}{N_o}}, \quad (13)$$

where $\sigma$ is the per-point uncertainty we defined in Equation (1), $N_i$ is the number of data points in-transit during the spot occultation, and $N_o$ is the number of data points in-transit but out-of-spot-occultation.

If we know the spot parameters accurately, we can calculate $\Delta D_{\mathrm{spot}}$ from Equation (10). However, an observer will likely have to estimate $\Delta D_{\mathrm{spot}}$ and $N_i$ directly from the lightcurve. Additionally, as mentioned at the start of Section 3.3, our injected spot models are generated using `starry`, which smooths the stellar surface, altering the shape of SCEs compared to a "hard-edged circle" model, such as in Equation (10). Therefore, to calculate $\Delta D_{\mathrm{spot}}$ we used `starry` to model the same star–planet system without spots. We then divided the `starry` model with the spot by the lightcurve for the no-spot system, rescaled to the contaminated transit depth, leaving only the spot bump (and small ringing artifacts). We calculated $N_i$ as the number of exposures within the full width at half-maximum (FWHM) of the SCE. We chose to use the FWHM as an easily calculable metric that is robust to the cadence, asymmetries, and exact start and end of the SCE (that may be lost in noise). Though by using the FWHM we will slightly underestimate $N_i$ and overestimate $N_o$, we expect this effect to be minor. $\Delta D_{\mathrm{spot}}$ is then taken as half the maximum height of this bump, to be consistent with using the FHWM.

## 4. Injection Recovery

### 4.1. Starspot Parameter Retrieval

We created 1000 SCEs using the method and parameter distributions described in Section 3.2. The parameters and their priors are outlined in Tables 1, 2, and 3. To generalize our fitting method for all spot crossings, we assume large, uninformative priors for the spot, allowing it to exist anywhere on the surface, including entirely behind the observed stellar face, allowing the model to "hide" spots if they are not justified. In total, for each spot crossing we fit for 12 parameters: stellar radius $R_*$, quadratic limb-darkening coefficients ($u_1$, $u_2$), spot contrast $C$, spot radius $R_{\mathrm{spot}}$, spot latitude $\phi_{\mathrm{spot}}$, spot longitude $\lambda_{\mathrm{spot}}$, planetary radius $R_{\mathrm{p}}$, planetary inclination $i$, transit epoch $t_0$, $p_0$, which acts as a constant scaling factor, and $n_\sigma$, an uncertainty inflation factor. We only fit for dark spots with positive contrasts, $C > 0$. We elected to fix the stellar mass to avoid degeneracy with stellar radius and to fix the orbital period because we are only considering a single transit with limited periodic information. We also note that while we fix the planetary inclination $i = 0$ for injection, we do not constrain $i$ when fitting, allowing the planet to transit away from the stellar equator to capture additional degeneracy with the spot location and limb-darkening. We used 30° of spherical harmonics in `starry` to model each system and limited the amount of smoothing over the surface map by setting the *spot_smoothing* parameter (standard deviation of Gaussian smoothing) in `starry` to 1/30. If the smoothing parameter is too low, we retain some ringing artifacts, and if it is too high, it can

**Table 1**
The Stellar Parameters of the `chromatic_fitting` Model Along with the Corresponding Prior Distributions

| Parameter | Units | Fix/Fit | Prior |
|---|---|---|---|
| $R_*$ | $R_\odot$ | Fit | $\mathcal{N}(1.0,\ 0.2)$, $R_* > 0$ |
| $M_*$ | $M_\odot$ | Fix | 1.0 |
| $P_{\mathrm{rot}}$ | d | Fix | 1000 |
| ($u_1$, $u_2$) | ... | Fit | D. M. Kipping (2013) parameterization |
| $i_*$ | ° | Fix | 90 |
| $C$ | ... | Fit | $\mathcal{U}(0.0, 1.0)$ |
| $R_{\mathrm{spot}}$ | ° | Fit | $\mathcal{U}(5, 45)$ |
| $\phi_{\mathrm{spot}}$ | ° | Fit | $\mathcal{U}(-135, 0)$ |
| $\lambda_{\mathrm{spot}}$ | ° | Fit | $\mathcal{U}(-135, 135)$ |

**Note.** Some parameters are kept constant during the fitting, these are marked with "fix" in the above tables and their respective priors are those fixed values.

**Table 2**
Same as Table 1 but for Planetary Parameters

| Parameter | Units | Fix/Fit | Prior |
|---|---|---|---|
| $M_{\mathrm{p}}$ | $M_\odot$ | Fix | 0.0 |
| $R_{\mathrm{p}}$ | $R_\odot$ | Fit | $\mathcal{N}(0.10, 0.02)$, $R_{\mathrm{p}} > 0$ |
| $i$ | ° | Fit | $\mathcal{N}(90, 0.1)$ |
| $P$ | d | Fix | 1.0 |
| $e$ | ... | Fix | 0 |
| $t0$ | d | Fit | $\mathcal{N}(0.0, 0.1)$ |

**Table 3**
Same as Table 1 but for Other Model Parameters

| Parameter | Units | Fix/Fit | Prior |
|---|---|---|---|
| $p_0$ | ⋯ | Fit | $\mathcal{N}(1.0, 0.1)$ |
| $n_\sigma$ | ⋯ | Fit | $\mathcal{N}(1.0, 0.005)$, $1 \leqslant n_\sigma \leqslant 3$ |

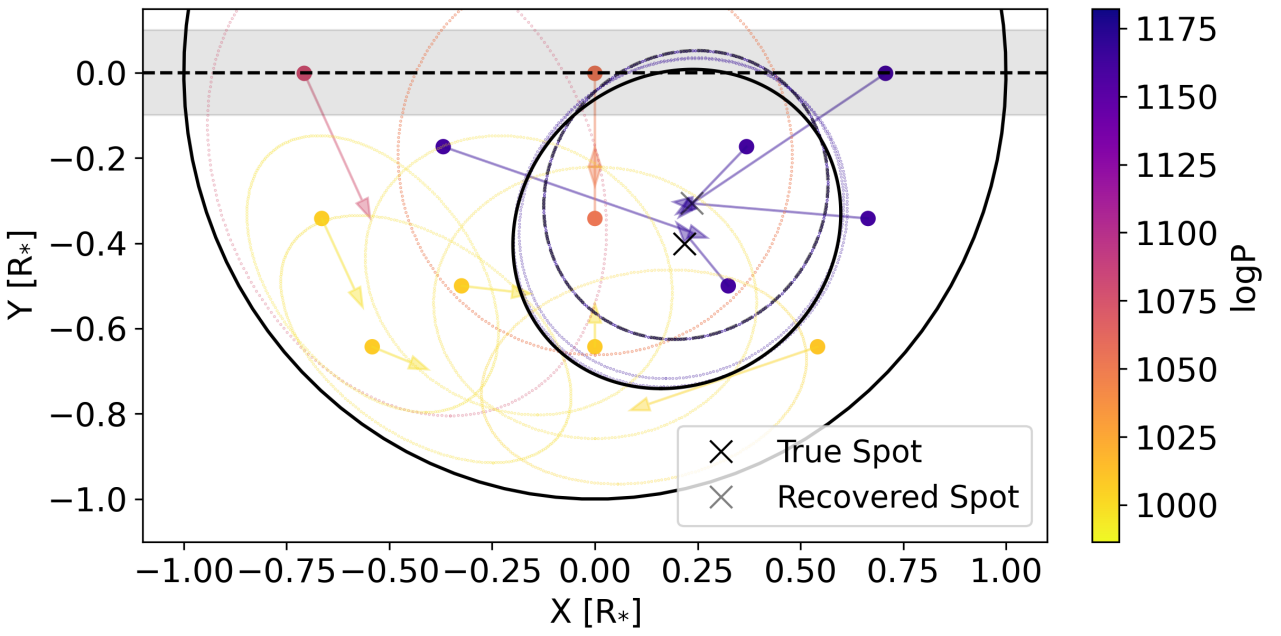

**Figure 3.** For the fourth spot-crossing scenario in Figure 1 (SNR = 11.5) we plot the log-likelihood values (colorbar) for the MAP-optimized transit and spot model for 13 different starting positions (circles). The final optimized spot locations for each starting position is indicated with an arrow. The true model is plotted with a black circle (a cross marking its center) and the highest likelihood spot is shown with a dotted gray circle. The transit chord is marked by the shaded gray region along the star's equator ($b = 0$).

dampen the spot contrast.[4] With this smoothing parameter and number of spherical harmonics we find that the minimum spot radius for which we can obtain <10% error in the contrast is ∼5°. Therefore, we decided not to inject or recover any spot smaller than 5°. We find that with orders of spherical degrees above 30 we run into numerical instabilities.

With 30° of spherical harmonics it was not computationally feasible to perform full MCMC sampling for all 1000 events. Instead, for this large population, we performed maximum a priori (MAP) optimization to retrieve the best-fit spot parameters and chose a handful of cases to MCMC sample in Section 5. The initial values for longitude and latitude were found to strongly affect the quality of the optimization. Spot degeneracies result in a heavily multimodal posterior space, which means it is easy for optimization and sampling methods that rely on local gradients to get "stuck" in local maxima, and not explore the full parameter space (D. Foreman-Mackey et al. 2013; D. W. Hogg & D. Foreman-Mackey 2018; see Section 6.3 of J. Dunkley et al. 2005). Therefore, for each SCE, we opted to loop the optimization over a series of 13 starting positions, choosing the one that produced the largest log-likelihood value (shown in Figure 3).

The results of the injection-recovery tests are shown in Figures 4 and 5 and Table 4. To discuss our results, we separate 1000 spot crossings into (i) low SNR (<4, $n = 306$) and (ii) high SNR (⩾4, $n = 694$). In Figure 4, we see a comparison between the injected and recovered spot parameters (contrast, radius, latitude, and longitude) and transit depth. By comparing Figures 4(a) and (b), we conclude that our recovery method fails to recover any parameter accurately for low SNR (i). Interestingly, the recoveries of most parameters are not significantly degraded by whether the spot overlaps with the limbs (any spot that touches the stellar limb is marked with a cross in Figure 4), except for spot longitudes. The median difference between injected and recovered, $|\Delta\lambda_{\rm spot}|$, for limb-spot cases is (i) $|\Delta\lambda_{\rm spot}| = 21.^\circ2$, ($n = 137$) and (ii) $|\Delta\lambda_{\rm spot}| = 0.^\circ75$ ($n = 244$), notably worse than the results in Table 4. However, for the other spot parameters ($C$, $R_{\rm spot}$, and $\phi_{\rm spot}$), there is no significant difference in the recoveries for limb and nonlimb spots. Here, we have considered all spots that appear to touch the stellar limbs as "limb spots"; however, how much of the spot projection is on or over the stellar limb may correlate with the quality of parameter recovery, but we will not explore this effect in this paper. We can see clear examples of poorly recovered limb spots in Figure 6.

From this point onward we only consider (ii), or spots with SNR ⩾ 4, as clear SCE detections. In Figure 5, we present the recovered spot and planet parameters as functions of percentile. A percentile of 80% means that 80% of the samples are recovered to within that y-value (e.g., 80% of spot radii are recovered to within 5°, or within 19%, of the injected values). We present the median absolute (top) and fractional (bottom) differences between the injected and recovered parameters in Table 4.

For the spot's position on the stellar surface, we recover an average difference of $|\Delta\lambda_{\rm spot}| = 0.^\circ15$ and $|\Delta\phi_{\rm spot}| = 3.^\circ13$. 80% of $\lambda_{\rm spot}$ and $\phi_{\rm spot}$ are recovered to within $1.^\circ0$ and $8.^\circ7$, respectively. We see similar results when plotting the injected and recovered spot positions in Figure 6, which shows a strong recovery of longitude (analogous with $x$-position) and a poorer recovery of latitude (analogous with $y$-position). The mean and median projected distances between injected and recovered positions are 0.99 and 0.64 $R_{\rm p}$, respectively.

For the spot's contrast and radius, we recover an average difference of $\Delta C = 4\%$ (or $|\Delta C| = 0.02$) and $\Delta R_{\rm spot} = 6.6\%$ (or $|\Delta R_{\rm spot}| = 1.^\circ3$). Meanwhile, 80% of spot contrasts and radii are recovered to within 17% (or $|\Delta C| = 0.15$) and 19% (or $|\Delta R_{\rm spot}| = 5.^\circ0$) of the injected values. Due to the strong degeneracies between latitude, radius, and contrast, we would expect challenges when recovering these parameters.

We recover the transit depth ($D_{\rm true} = 10{,}000$ ppm) of the planet with an average difference, $\Delta D$ of 78.3 ppm, or 0.78%. In 80% (90%, 95%) of cases we recover $D$ to within 253 ppm (432 ppm, 640 ppm), or 0.6% (1.8%, 3.5%).

### *4.2. Retrieval of Transit Parameters When Fitting versus Masking a Spot*

If, instead of fitting for the spot, we were to mask each spot crossing, we would recover transit depths contaminated by the TLSE (see Equation (9)). With multiwavelength observations and sufficient coverage in the visible (where spot contrasts are highest), we could then correct for this contamination directly in transmission spectra; however, with the major drawbacks discussed in Section 3.1.

In Figure 7, we demonstrate the improvement of the recovered transit depths for our spot-crossing sample (with SNR ⩾ 4) compared to the contaminated transit depths we would recover with masking. By fitting spot crossings we recover the true transit depth to within 1% (2%, 5%) in 70% (84%, 93%) of samples. From Figure 7, we see the greatest improvement in the recovered depth for large contaminations, and we can overcorrect for the spot and retrieve depths smaller than the injected value toward smaller contaminations. For $\epsilon \geqslant 1.013$ (1.027, 1.087), we retrieve a depth closer than masking to the true value in 95% (98%, 100%) of injection-recoveries. As $\epsilon$ is a difficult parameter to visualize, these values of $\epsilon$ are demonstrated in Figure 8, showing the relationship with the more intuitive projected spot-coverage fraction, $f$, and spot contrast, $C$. From Figure 8, when spots have $C < 5\%$ (which we do not explicitly test in this work) or $f < 2\%$[5]

[4] For more information on the spot smoothing parameter in `starry` see the documentation here: https://starry.readthedocs.io/en/latest/notebooks/StarSpots/.

[5] Since we define $f$ as the projected spot-covering fraction of the visible hemisphere, $f < 2\%$ would be equivalent to $R_{\rm spot} < 0.014 R_*$ if the spot was perfectly in the center of the star ($\lambda_{\rm spot} = \phi_{\rm spot} = 0°$). However, the projection effects of moving the spot off-center decrease $f$, and therefore the equivalent constraint on $R_{\rm spot}$ could be lower.

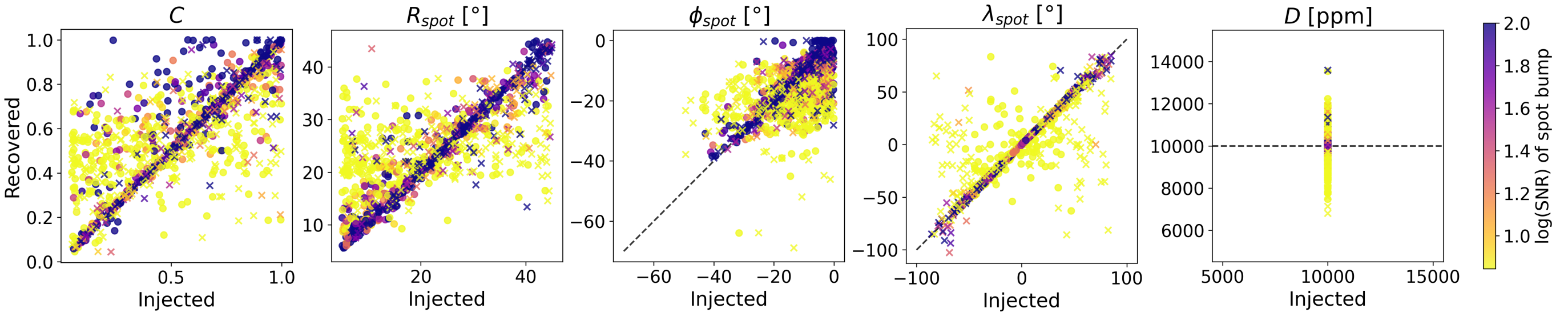

**(a)** All (SNR≥0)

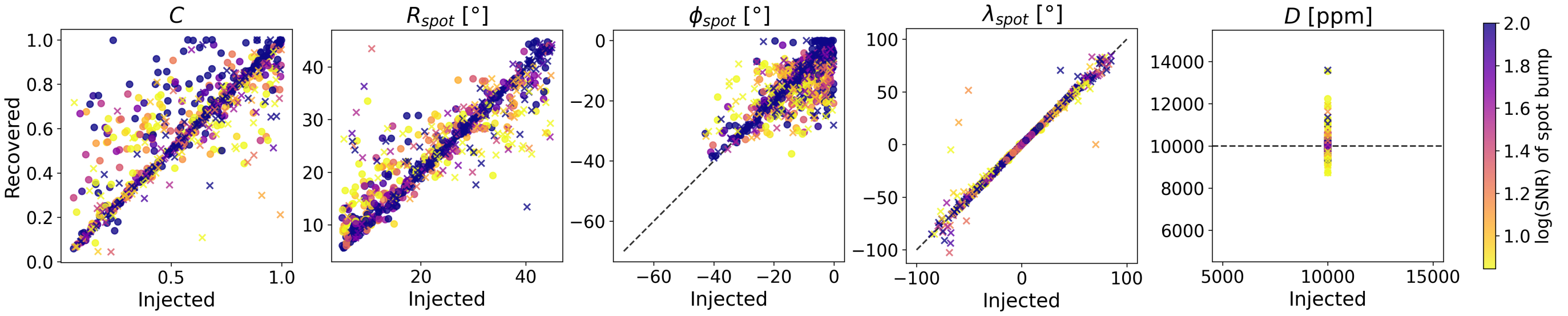

**(b)** SNR≥4

**Figure 4.** Recovered versus injected spot contrast, radius, latitude, longitude, and transit depth for (a) all SCEs and (b) SCEs with SNR $\geqslant$ 4. The points marked with crosses are spots whose projections overlap with the stellar limb. Limb spots are much more difficult to fit and have intrinsically much more uncertainty. The colorbar represents the log(SNR) of each SCE.

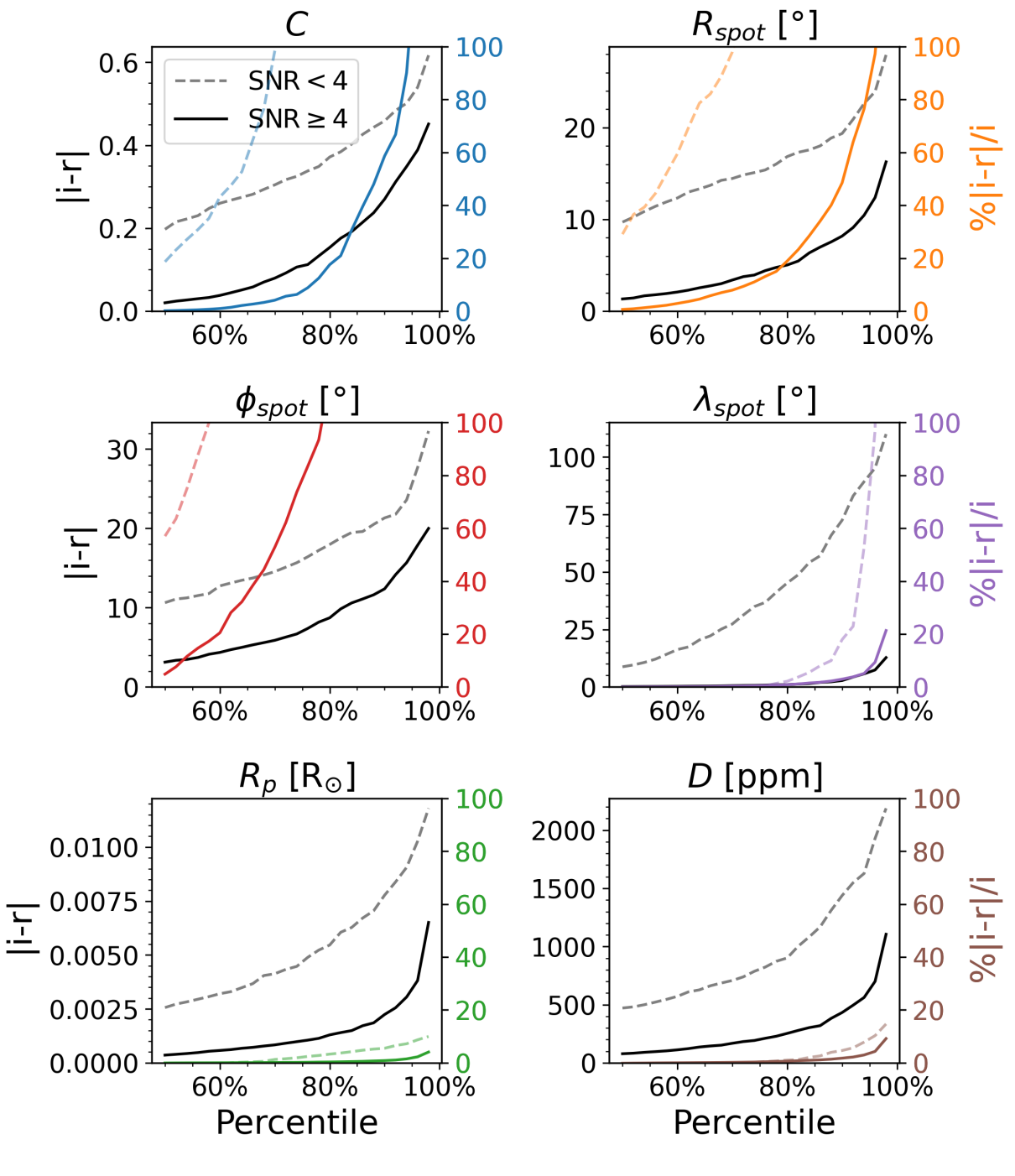

**Figure 5.** For SNR<4 (dashed lines) and $\geqslant$4 (solid lines), we show five recovered parameters: spot contrast (top left), radius (top right), latitude (middle left), longitude (middle right), and planetary radius (bottom left). On the left axis (black), we plot the percentile of the absolute difference between injected, $i$, and recovered, $r$. On the right axis (colors), we show the percentage difference from the injected value, also against percentile. Only percentiles from 50% to 98% are shown here.

over 50% of transit depth recoveries are worsened by fitting. Therefore, we can conclude that these small or low contrast spots are likely better masked, not fitted. We acknowledge that this is a circular problem—here we must model the SCE first to derive $\epsilon/C/f$. However, if we have additional observations of the system, such as the rotational photometric variability, this could provide some constraint on the spot parameters without fitting the SCE. However, if the spot does need to be modeled to derive $\epsilon$, and $\epsilon$ is found to be very low, it could be beneficial to mask the SCE and assess how the difference in recovered transit depths between masking and fitting affects inferences made about the planet or its atmosphere.

**Table 4**
Median Percentage and Absolute Differences between Injection, i, and Recoveries, r, of Spot-crossing and Planetary Parameters: Spot Contrast, $C$, Spot Radius, $R_{\rm spot}$, Spot Latitude, $\phi_{\rm spot}$, Spot Longitude, $\lambda_{\rm spot}$ and Planetary Radius, $R_{\rm p}$ We Present the Results for Low and High SNR Scenarios

| | Median % $\|i\text{-}r\|/i$ | |
|---|---|---|
| | SNR < 4 | SNR $\geqslant$ 4 |
| $\Delta C$ [%] | 42.7 | 4.0 |
| $\Delta R_{\rm spot}$ [%] | 43.7 | 6.6 |
| $\Delta \phi_{\rm spot}$ [%] | 59.1 | 31.6 |
| $\Delta \lambda_{\rm spot}$ [%] | 39.5 | 0.78 |
| $\Delta R_p$ [%] | 2.56 | 0.36 |
| $\Delta D$[%] | 4.7 | 0.78 |
| | Median $\|i\text{-}r\|$ | |
| | SNR<4 | SNR $\geqslant$ 4 |
| $\|\Delta C\|$ | 0.20 | 0.020 |
| $\|\Delta R_{\rm spot}\|$ [°] | 9.7 | 1.3 |
| $\|\Delta \phi_{\rm spot}\|$ [°] | 10.6 | 3.1 |
| $\|\Delta \lambda_{\rm spot}\|$ [°] | 8.8 | 0.15 |
| $\|\Delta R_p\|$ [$R_*$] | 0.00256 | 0.00036 |
| $\|\Delta D\|$ [ppm] | 471.2 | 78.3 |

We visually inspected the five scenarios with the largest difference between the recovered depth and the true depth (i.e., the biggest outliers). The spots in these scenarios all had extremely large, nonphysical contrasts ($C = 0.98, 0.99, 0.88, 0.95, 0.96$) and radii ($R_{\rm spot} = 44°, 45°, 41°, 30°, 42°$) and four of the five were located on the stellar limb, returning the start or end of the transit duration to baseline flux. These five largest outliers are shown in Figure A1 in the Appendix. Since these spots are unphysical, we are not concerned about these cases.

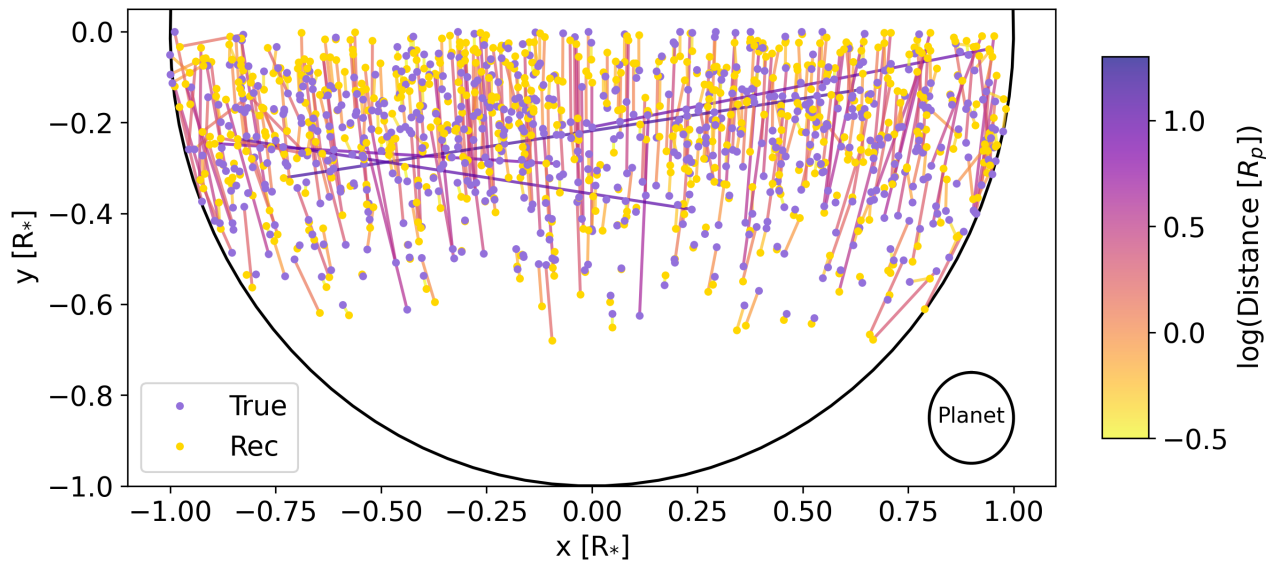

**Figure 6.** The injected (purple) and recovered (gold) center positions of the 694 SCEs with SNR $\geqslant 4$. A line is plotted joining the two positions, with its color determined by the distance between injected and recovered (in planetary radii, $R_p$). The spots that occur on or near the stellar limb have the largest differences between the injected and recovered positions. The planet size is plotted in the lower right corner for scale.

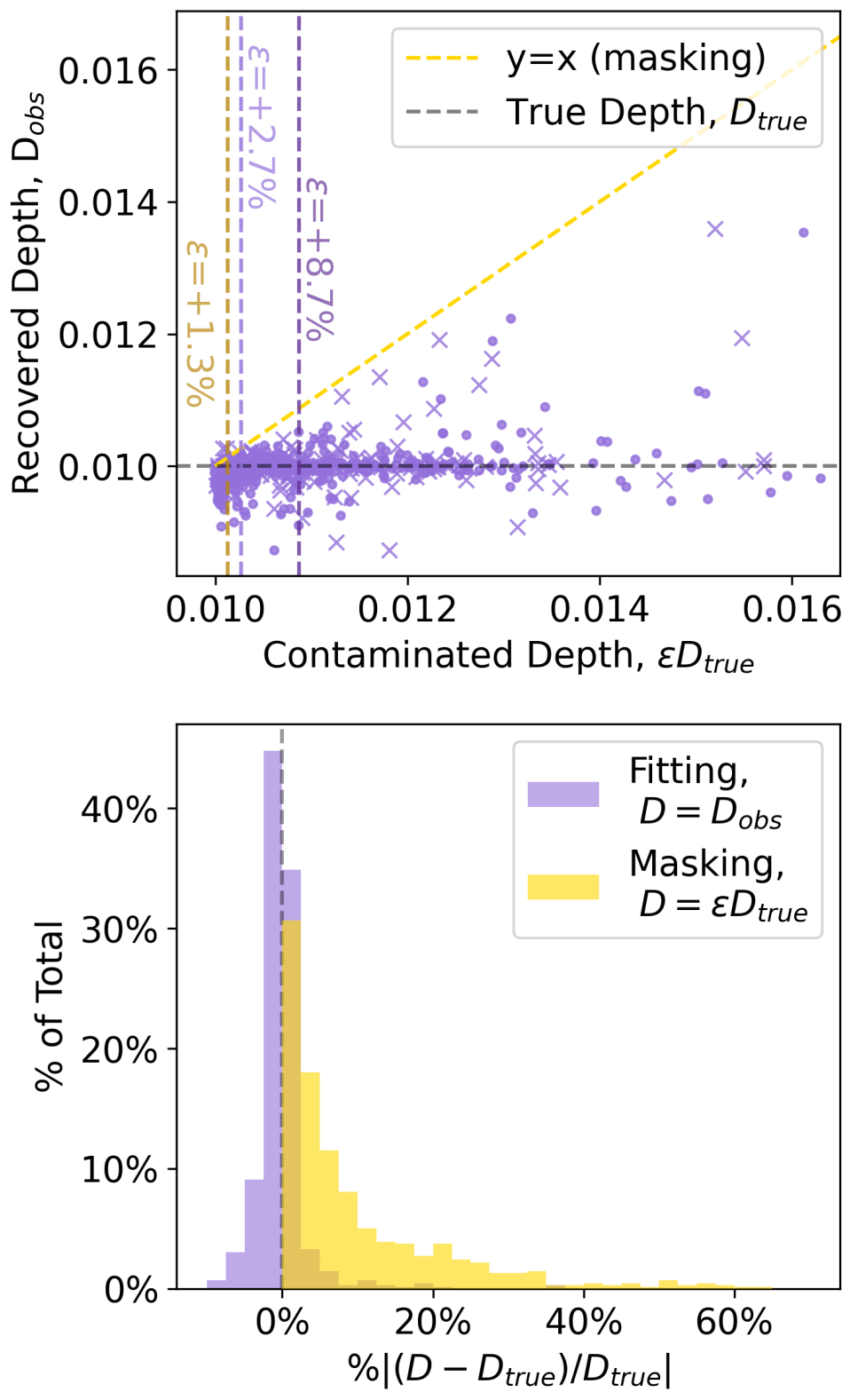

**Figure 7.** A comparison of the recovered transit depths, $D$, from fitting the spot crossing versus masking. In the masking case we assume that we do not have any prior information on the stellar contamination and so the recovered depth would simply be the contaminated depth $\epsilon D_{\rm true}$. Upper: The recovered depths from fitting, $D_{\rm obs}$, against the injected contaminated depths. Spots on the limb are marked by crosses. The dotted gold line shows the $y = x$ line (indicating the depths recovered by masking the spot) and the dotted gray line is the true depth injected, $D_{\rm true}$. We include three notable $\epsilon$ values as vertical dotted lines (1.3% in dark gold, 2.7% in purple, and 8.7% in dark purple). Lower*:* A histogram of the percentage differences between $D_{\rm true}$ and $D_{\rm obs}$ (purple) and $D_{\rm true}$ and $\epsilon D_{\rm true}$ (orange).

Overcorrection of the transit depths for small contamination signals is likely driven by strong degeneracies between spot parameters. While all spot crossings yield degenerate solutions, it becomes more pronounced when the injected occulted spot is very small (comparable to the planet radius) and/or low in contrast. In such cases, the optimization tends to favor degenerate solutions with larger sizes or higher contrasts, especially near the prior limits. This effect is likely exacerbated by imposing a lower bound on the spot radius (>5°). We demonstrate this in Figure A2 in the Appendix, which shows the five scenarios with the most overcorrected (i.e., smallest) recovered transit depths. In three of these cases, the injected spot size approaches the lower limit of 5° and all injected spot contrasts are $\leqslant 0.24$. In all five scenarios, the injected and recovered lightcurves are visually indistinguishable, highlighting how parameter degeneracies can obscure accurate spot recovery.

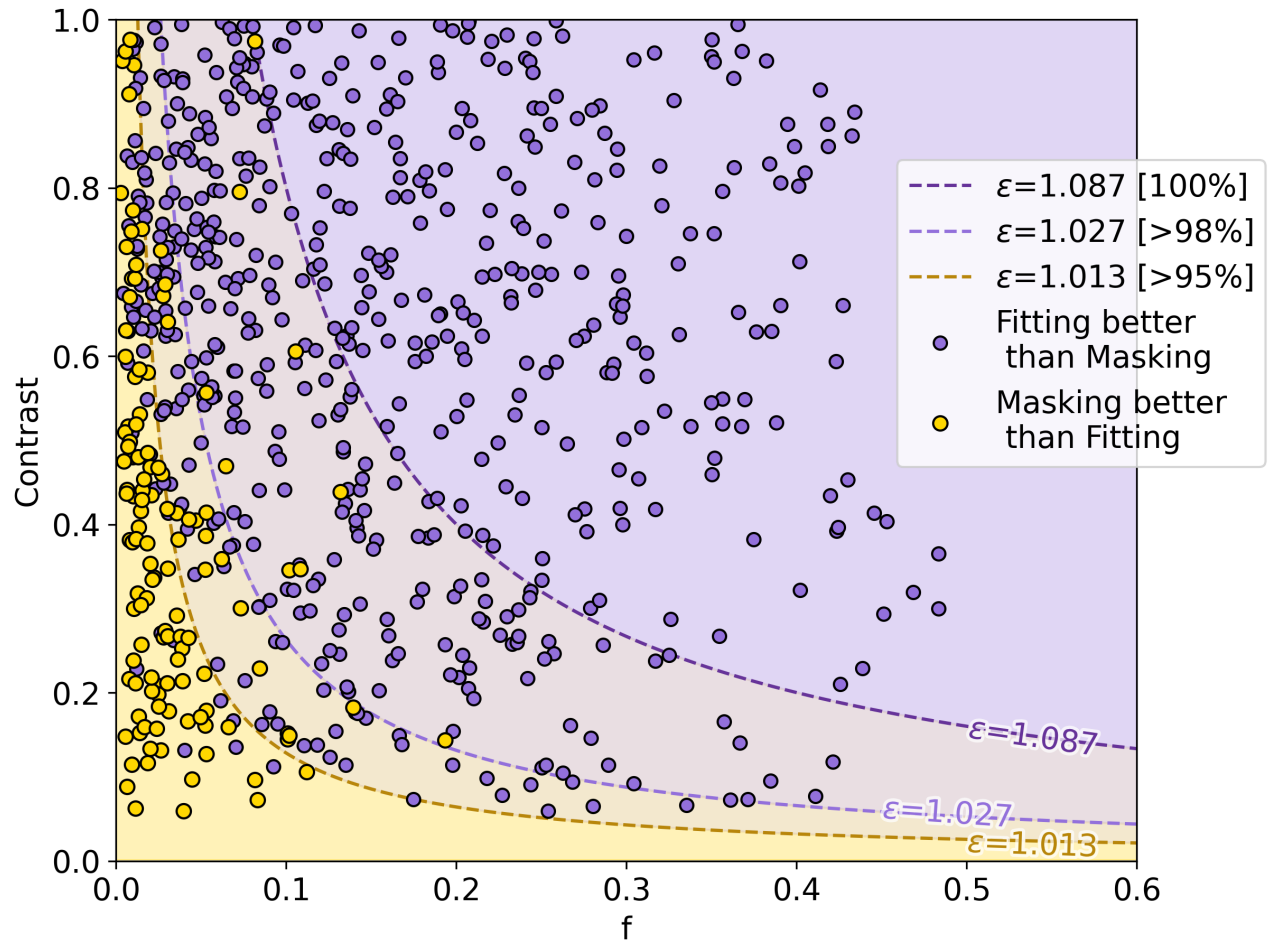

**Figure 8.** The spot contrast and covering fraction, $f$, for all SCEs with SNR $\geqslant 4$. Scenarios where fitting for SCEs improves the transit depth recovery, compared to masking, are marked in purple and vice versa in gold. The dotted lines correspond to $\epsilon = 1.087$ (dark purple), 1.027 (medium purple), and 1.013 (dark gold). When $\epsilon > 1.087$ [1.027, 1.013], fitting improves recovery of the transit depth in 100% [98%, 95%] of cases and for $\epsilon < 1.087$ [1.027, 1.013], it improves 59% [50%, 26%].

### *4.3. The Presence of Unocculted Spots*

Our spot-modeling approach has thus far only taken into account one spot. In the (likely) case that there are additional unocculted spots on the stellar surface, we have only partially mitigated the TLSE contamination in the transmission spectrum. Therefore, we have to further correct the transit depth for the additional spot-covering fraction we have not yet considered. We can extract the spot contrast, $C_\lambda$, and the covering fraction of one spot, $f_1$, from our spot-crossing model and fit for the total spot-covering fraction, $f_{\rm total}$ (which we know is $=f_1 + f_{\rm unocculted}$), using the transmission spectrum, through the following equation:

$$D_{\rm true} = \left(\frac{1 - f_{\rm total}\, C_\lambda}{1 - f_1\, C_\lambda}\right) D_{1,\lambda}, \tag{14}$$

$$D_{\rm true} = \frac{\epsilon_{1,\lambda}}{\epsilon_\lambda} D_{1,\lambda}, \tag{15}$$

where $D_{1,\lambda}$ is our transmission spectrum corrected for only one spot, $\epsilon_\lambda = 1/(1\text{-}f_{\rm total} C_\lambda)$ is the contamination factor from B. V. Rackham et al. (2018), and $\epsilon_{1,\lambda}$ is, equivalently, the contamination from one spot only. Assuming our SCE modeling is accurate, and that all spots on a star share a common temperature, we have already derived the spot contrast; therefore, this method involves only fitting for the remaining spot-covering fraction. The key advantage of this approach is that it bypasses fitting stellar models to derive the spot contrast—thereby avoiding the model fidelity issues discussed earlier.

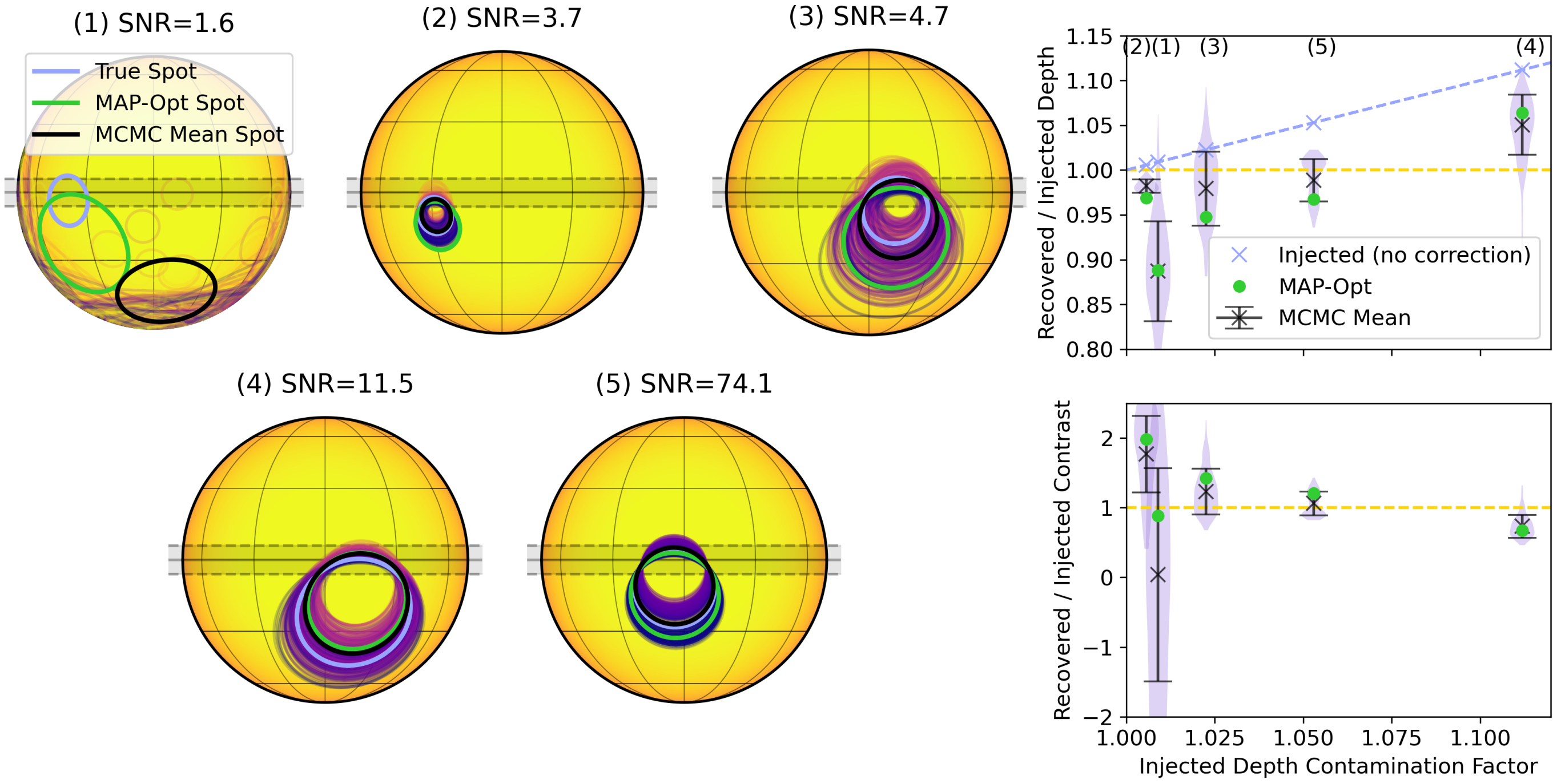

**Figure 9.** Five star plots: The five injection-recovery scenarios shown in Figure 1, with SNR = 1.6, 3.7, 4.7, 11.5, and 74.1, respectively. For each of 300 random MCMC traces, we plot a semitransparent representation of the spot. The color of each trace is determined by the spot's contrast with the star. The planet's transit across the star is shaded in light gray. The injected (or true) spot in each case is plotted in purple. The MAP-optimized spots are in green and the MCMC mean spots are in black. Right top: The recovered/injected transit depths as a function of the contaminated depth factor, $\epsilon$, from the TLSE. The dashed yellow line is where the recovered = injected transit depth (i.e., the spot is perfectly corrected) and the dashed light blue line is where recovered = contaminated transit depth (i.e., the spot is not corrected at all, equivalent to masking). For each of the SCEs, we add violin plots for the traces. Right bottom: The recovered/injected spot contrasts as a function of $\epsilon$.

## 5. MCMC Sampling

We chose to perform full MCMC sampling on five representative samples (the examples shown in Figure 1) with a range of spot SNRs, sizes, and locations. MCMC sampling provides an approximation of the posterior distribution, allowing us to directly extract parameter uncertainties and explore covariances. In theory, MCMC sampling is a great way to explore degeneracies in parameter space; however, in reality, sampling of multimodal posterior space is highly sensitive to the chosen initial values and chains can get "stuck" at local likelihood maxima (J. Dunkley et al. 2005). We at least partially mitigate this issue by starting from the highest likelihood spot location from MAP optimization in each case.

To MCMC sample we used the `chromatic_fitting` models for our chosen starspot injection scenarios. `chromatic_fitting` uses `PyMC3` (J. Salvatier et al. 2016) as its MCMC sampling framework. In this work we exclusively use the No U-Turn (NUTS, M. D. Hoffman & A. Gelman 2014) sampler. We reduced the number of spherical harmonic orders `starry` uses to create the stellar map to 26 (compared to 30 for the injection and MAP optimization) to speed up the MCMC sampling. Higher orders of spherical harmonics allow for a greater surface resolution; however, it is extremely time intensive. With 26° we should have <10% errors on the contrast down to spot radii of 5°.[6] We also increased the uniform prior for spot contrast from 0–1 to −1–1 to include the possibility of "hot" spots and to better explore the posterior space for badly constrained spots (e.g., where SNR < 4). For each scenario, we ran 1000 tuning steps and 3000 draws across 2 chains. Usually, we ensure that the chains have fully converged using the Gelman–Rubin statistic (A. Gelman & D. B. Rubin 1992). However, for these scenarios with many degenerate solutions the degenerate parameters (spot parameters and planet radius) often have a Gelman–Rubin statistic >1.1, while other parameters converge well. For these parameters, we may not be fully exploring the complete posterior distribution effectively, and our uncertainties for these parameters should be taken as underestimates. In addition, since our aim here is to explore the degenerate space, we still consider MCMC solutions with larger Gelman–Rubin statistics than normal for the expected degenerate parameters.

In Figure 9, we show 300 random MCMC traces for each scenario. We can see that for the two lowest SNR (<4) cases, fitting for the spot returns a similar or worse transit depth than masking. In particular, for the lowest SNR = 1.6 case, the spot properties are entirely unconstrained. In all of the higher SNR > 4 cases (4.7, 11.5, 74.1) we improve the transit depth estimate by fitting compared to masking and we retrieve the true transit depth and spot contrast within error for two out of three; however, from the spread of MCMC traces it is clear that the recovery of the transit depths and contrasts are strongly affected by degeneracies.

[6] See https://starry.readthedocs.io/en/latest/notebooks/StarSpots/.

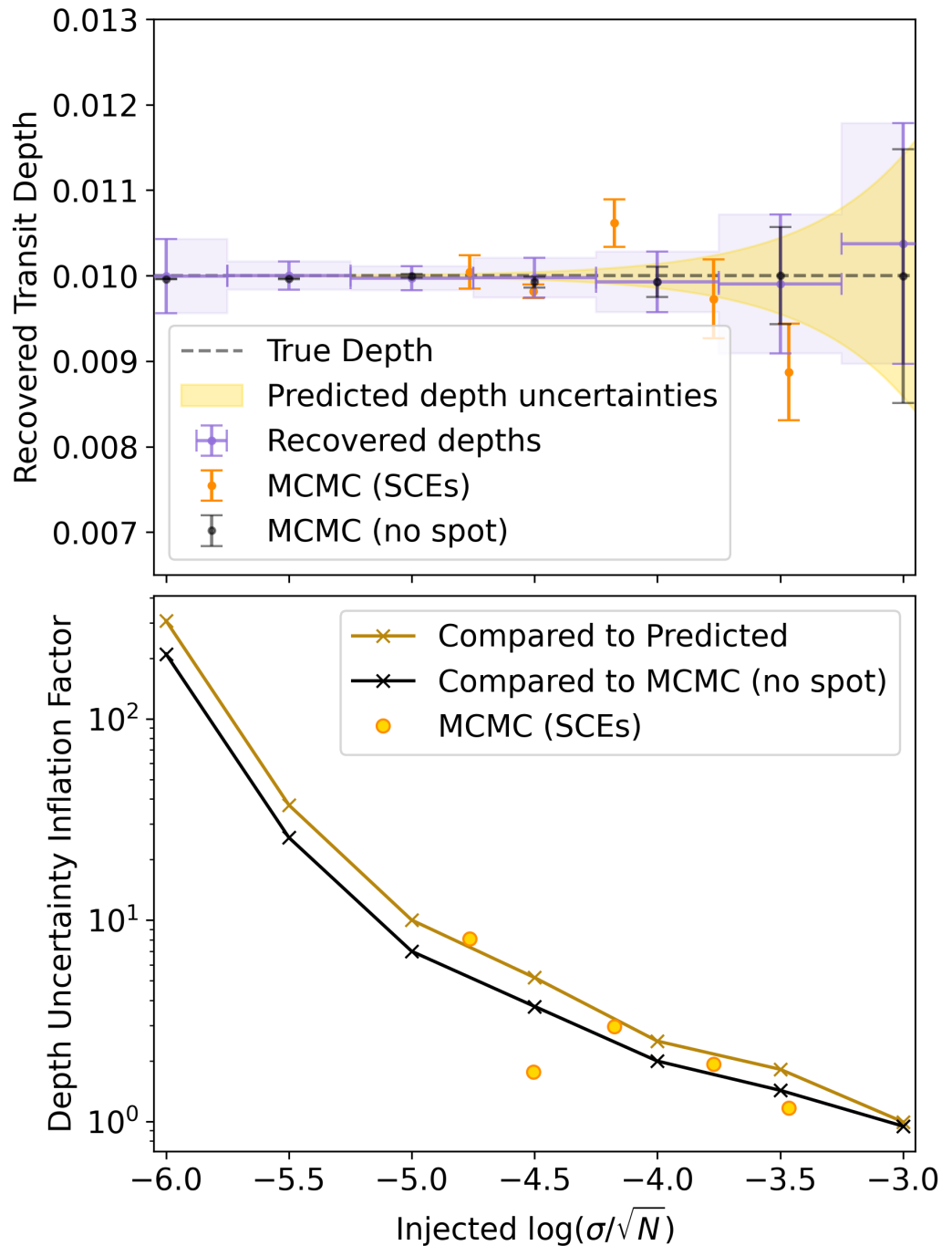

**Figure 10.** Top: Recovered transit depth as a function of injected per-point uncertainties divided by the square-root of the number of in-transit data points, $\sigma/\sqrt{N_{\rm transit}}$ (to remove dependence on cadence). The SCE injection-recovery results (Section 4, SNR > 4) are binned in $\log\sigma = 0.5$ steps; the median recovered transit depths and their standard deviations within each bin are shown in purple. Results from full MCMC sampling of the five SCEs in Figures 1 and 9 are in orange. The predicted transit depth uncertainties ($=\sigma\sqrt{2/N_{\rm transit}}$) are shown by the gold region, while MCMC-sampled results from unspotted lightcurves are shown in black. Bottom: Inflation of transit depth uncertainties in the presence of an SCE, relative to both the theoretical expectation (gold) and the unspotted MCMC baseline (black). Uncertainty inflation values for the five MCMC-sampled SCE cases are shown by orange points.

### 5.1. Uncertainty Inflation due to SCEs

To understand how SCEs affect uncertainties in the extracted transit depths, we compared our injection-recovery results to a baseline scenario in which the transit chord is free of stellar spots. In theory, the photon noise limit on the transit depth is given by $\sigma_{\rm depth} = \sigma\sqrt{2/N}$, where $\sigma$ is the per-point uncertainty, assuming equal numbers of exposures taken in and out of transit. However, in practice there are additional degeneracies in the transit fit between stellar, planet, and orbital parameters that complicate our retrievals and inflate depth uncertainties. Using the same model setup described in Section 4.1, but with no spots on the stellar surface, we generated seven lightcurves. Each lightcurve had injected Gaussian noise with standard deviations ranging from $10^{-5}$ to $10^{-2}$ in logarithmic increments of $10^{0.5}$. We then performed MCMC sampling (4 walkers, 5000 tuning steps, 10,000 draws) on each lightcurve, using the stellar and planetary priors listed in Tables 1 and 2. The free parameters were the stellar and planet radii, quadratic limb-darkening coefficients, orbital inclination, transit epoch, and out-of-transit baseline. The resulting transit depths and their corresponding posterior uncertainties are shown in black in the top plot of Figure 10. The MCMC-fitted depth uncertainties range from 1.0 to 1.5× the theoretical uncertainty for $\sigma = 10^{-2}$–$10^{-5}$, respectively.

Next, we averaged the injection-recovery results (with SNR > 4) into the same uncertainty bins. Since the MAP optimization does not yield uncertainties directly, we approximated them by taking the standard deviation of the recovered transit depths within each bin, effectively marginalizing over spot size, location, and contrast. To quantify how transit depth uncertainties are inflated by SCEs, we divided the binned injection–recovery uncertainties by the theoretical photon noise limit (shown in gold in Figure 10) and the baseline MCMC results with no-spot crossings (shown in black).

Finally, we confirm that the five SCE scenarios, for which we performed full MCMC sampling (Section 5) and have better uncertainty estimates from posterior sampling, follow a consistent trend in uncertainty inflation as a function of $\sigma$, matching the behavior of the broader injection–recovery set.

We find that SCEs, even when fitted, inflate the uncertainties on recovered transit depths and that this inflation grows as $\sigma$ decreases. For JWST-level precisions ($\sigma \sim$ 10–100 ppm), we observe inflation factors ranging from few 10 to 100x the predicted values. This effect becomes more pronounced at lower $\sigma$ values, likely because high-precision lightcurves are more sensitive to subtle features introduced by starspot crossings, thereby amplifying the influence of spot degeneracies on the inferred parameters.

While the exact inflation factors will vary from case to case, these results highlight the need for caution when estimating transit depth uncertainties for stars where spot crossings are likely. This work is intended as a general road map. Ideally, for a given system, we would perform these injection-recovery tests on real or representative lightcurves, incorporating spot distributions informed by observations (rather than the generic ones defined in Equation (1)) to more accurately assess the impact of SCEs for that planet.

## 6. The Degeneracy Problem

As shown in Sections 4.1 and 5, deriving starspot properties from a single-band photometric lightcurve is limited by degeneracies that can inflate uncertainties and produce multi-modal posteriors. In this section, we provide a framework to anticipate and bound those degeneracies before fitting.

Spot latitude, contrast, and radius are strongly degenerate—especially when the stellar inclination is poorly constrained. For instance, assuming a 90° inclination, a small equatorial spot can produce a very similar photometric signature to a larger or darker spot at higher latitudes. These degeneracies have been extensively studied in the context of rotational modulation observed by missions such as Kepler and TESS (A. F. Lanza 2016; R. Luger et al. 2019; G. Basri & R. Shah 2020; B. M. Morris 2020). Transiting planets offer a unique advantage: they provide a detailed scan of the stellar surface along the transit chord. This spatial resolution can break degeneracies (A. F. Lanza 2016), especially when the planet occults an active feature such as a starspot (e.g., A. V. R. Silva 2003; B. M. Morris et al. 2017). When combined with long-term monitoring of stellar rotation, spot crossings during transit can help constrain the degenerate stellar surface considerably.

In this section, we demonstrate how three directly measurable observables of an SCE—its timing, duration, and amplitude—can be translated into geometric constraints on spot longitude, latitude, radius, and contrast. Unlike spot-modeling frameworks, such as `starry` or `spotrod`, which forward model spot parameters and rely on posterior samples to reveal degeneracy structure, we use geometric constraints to visualize the families of spot solutions that are consistent with these observables.

### 6.1. Observables and Their Link to the Spot Properties

In Sections 4 and 5, we used uninformative priors for the spot when optimizing and sampling (see Tables 1–3). However, an SCE has three distinct instantaneous parameters that we can leverage to inform our priors:

1. $t_{spot}$. The spot-crossing epoch or midpoint of spot crossing $t_{\rm spot}$ only correlates with the longitude of the spot center.
2. $\Delta t_{spot}$. The spot-crossing duration $\Delta t_{\rm spot}$ depends predominantly on the latitude and radius of the spot, though these two parameters are degenerate.
3. $\Delta D_{spot}$. The size of the bump in the lightcurve due to the spot $\Delta D_{\rm spot}$ also depends on the latitude and radius of the spot; however, apart from the transit depth, which may also depend on the planet's atmosphere, it is the only probe of spot contrast.

Together, these three observables define geometric constraints that carve out allowed regions in ($\phi_{\rm spot}$, $\lambda_{\rm spot}$, $R_{\rm spot}$, $C$) space.

Uncertainties in these parameters (i.e., uncertainties in time and flux) will then propagate into the derived spot parameters. The shape of the spot crossing, such as the skew and asymmetry, will also provide information about the geometry of the spot; however, this is more difficult to quantify and is considered when performing more detailed model fits to the lightcurve. Our aim in this section is not to model spots directly, as in previous sections but to use these three observables to constrain the stellar spot radius-position-contrast parameter space, generate useful priors for MCMC sampling, and better understand the degeneracies at play. This approach is conceptually similar to G. Basri & R. Shah (2020), who used lightcurve metrics related to the rotational variability of stars to derive starspot distributions and other useful physical information.

### 6.2. Exploring the Degeneracy Space

#### 6.2.1. Creating a Grid of Spot Parameters

To explore the degeneracy space and connect lightcurve observables to spot properties, we constructed a large 3D grid of spot parameters for the planet–star system defined in Section 3.2. The grid samples latitude ($\phi_{\rm spot}$), longitude ($\lambda_{\rm spot}$), and radius ($R_{\rm spot}$) with 1° resolution ranging from $\phi_{\rm spot} = -60°–0°$, $\lambda_{\rm spot}$ = –135°–0°, and $R_{\rm spot}$ = 1°–45°, rendering a total of 373,320 grid points. Since we have assumed an equatorial planet ($b = 0$), we do not consider $\phi_{\rm spot} < -60°$ because even the largest allowed spot ($R_{\rm spot} = 45°$) will not intersect with the transit chord. To reduce computation time we exploit the symmetry of our setup; only generating half the longitudes and latitudes and mirroring the results (to produce 1,475,595 grid points). Using the same time cadence as in Section 3.2, we model the planet and star using SHAPELY (as in Section 3.3.2). We model each spot as a hard-edged circle on the stellar surface, equivalent to a spherical cap, and we take into account the elliptical projection effects of moving the spot across the surface. Toward the limbs we only consider the part of the spot on the visible hemisphere ($z > 0$, i.e., faced toward us), effectively allowing the spot to "wrap" around the star.

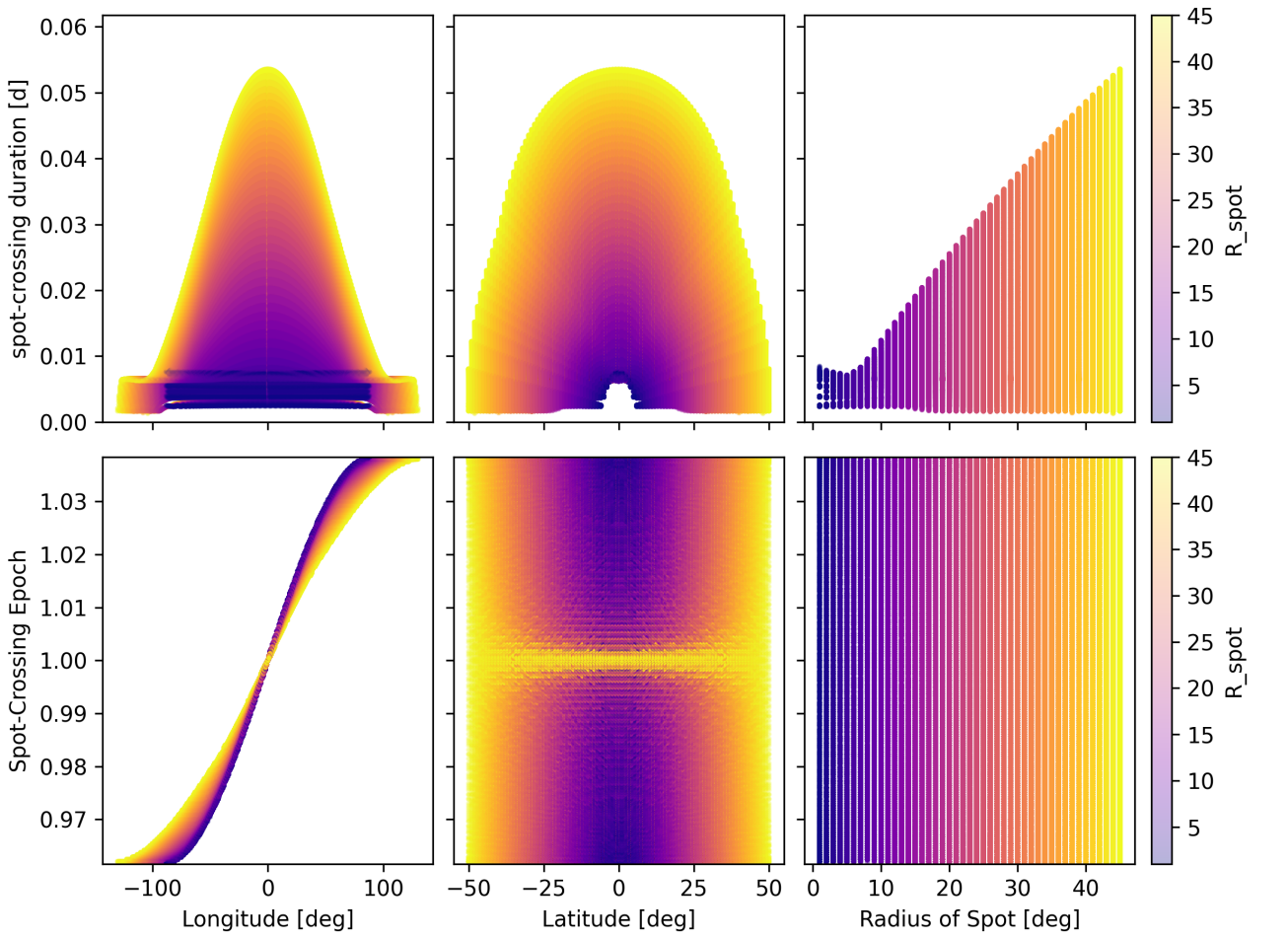

**Figure 11.** Spot-crossing duration, $\Delta t_{\rm spot}$, and spot-crossing epoch, $t_{\rm spot}$, plotted as functions of $\phi_{\rm spot}$, $\lambda_{\rm spot}$, and $R_{\rm spot}$ for the 1°-increment grid described in Section 6.2.1.

For each of the 1,475,595 latitude–longitude–radius combinations, we calculate:

1. Whether the spot is occulted by the planet during transit;
2. The total projected fractional spot-coverage of the visible stellar disk, $f$;
3. The instantaneous spot-covered fraction of the planet's shadow, $g(t)$, following Equation (9).

This enables us to extract two key observables for each lightcurve: the spot-crossing duration ($\Delta t_{\rm spot}$), defined as the FWHM of the spot crossing as in Section 3.3.3, and the spot-crossing midpoint ($t_{\rm spot}$). We remove any grid points where the spot is not occulted. These quantities are then plotted as functions of $\phi_{\rm spot}$, $\lambda_{\rm spot}$, and $R_{\rm spot}$ in Figure 11.

From Figure 11, it is clear that the value of $t_{\rm spot}$ tightly constrains $\lambda_{\rm spot}$ but does not reduce the $\phi_{\rm spot}$ or $R_{\rm spot}$ parameter space. $\Delta t_{\rm spot}$ places a minimum bound on spot size, or, alternatively, the spot radius sets an upper limit on the possible spot-crossing duration. $\Delta t_{\rm spot}$ also sets a loose range on longitude (as the spot begins to wrap around the stellar limbs, for larger spots $\Delta t_{\rm spot}$ quickly decreases) and latitude (through the banana-shaped latitude-radius degeneracy seen later in Figure 13(e)). Both these ranges are strongly dependent on the maximum spot size set; if we instead impose $R_{\rm spot} < 20°$, the same duration will constrain to smaller possible ranges for $\lambda_{\rm spot}$ and $\phi_{\rm spot}$. We note that there are a couple of apparent artifacts present in Figure 11 when $\Delta t_{\rm spot} < 0.008$ day ≈ planet-crossing time. First, we see that $\Delta t_{\rm spot}$ starts to increase for $R_{\rm spot} < 5°$, likely because this is approximately the planet radius $R_{\rm p}$ and we have different SCE shapes for grazing spots (smooth bumps) and fully eclipsed spots (flat-topped), and therefore different FWHMs. At these small durations we also see a plateau in longitude (where $|\lambda_{\rm spot}| \gtrsim 100°$) and a small gap for latitude near the stellar equator. Longitude is unconstrained below this duration because all spots smaller than the planet ($R_{\rm spot} \lesssim R_{\rm p}$) will produce a similar duration. Similarly, the latitude gap happens because when the spot is contained within the transit chord the planet must transit at least $R_{\rm spot}$ (at the limbs, $2R_{\rm spot}$ at $\lambda_{\rm spot} = 0$), setting a minimum duration for these spot crossing that correspond to min($R_{\rm p}$, $R_{\rm spot}$) ≈ 5°.

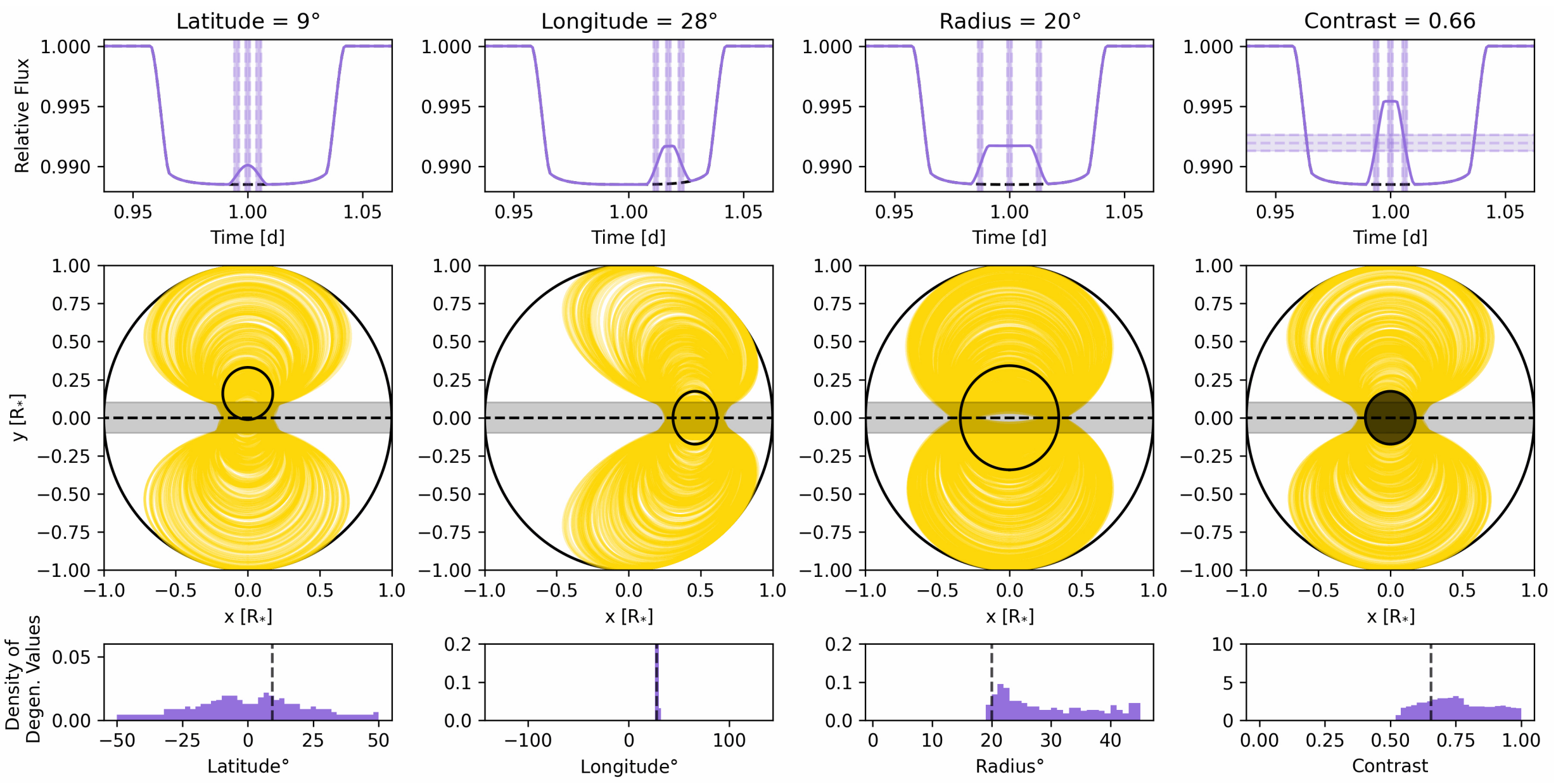

**Figure 12.** Animation demonstrating how the degeneracy region changes as we step through spot $\phi_{\rm spot}$ (left), $\lambda_{\rm spot}$ (left middle), $R_{\rm spot}$ (right middle), and $C$ (right), while keeping the other three parameters fixed in each case. Top: The base transit lightcurve in dotted black and the lightcurve with spot crossing in purple. The resulting $\Delta t_{\rm spot}$ and $t_{\rm spot}$ (and error regions) are indicated with vertical dotted purple lines. Middle: The star with the spot shown in black, the transit chord is shaded in gray and 500 degenerate solutions matching $\Delta t_{\rm spot} \pm \sigma_{\Delta t_{\rm spot}}$ and $t_{\rm spot} \pm \sigma_{t_{\rm spot}}$ are shown in gold. In the right plot, the shading of the spot represents its contrast. Bottom: Histogram of the spot parameters for the remaining degenerate solutions. The true value is shown by the black dotted line and in all cases the true parameter is contained within the histogram of solutions. As the animation steps through spot parameters, we see that the latitude is highly degenerate, longitude is tightly constrained except at the stellar limbs, and we can place a lower bound on the radius and contrast. The static image is a still from the animation at $\phi_{\rm spot} = 9°$, $\lambda_{\rm spot} = 28°$, $R_{\rm spot} = 20°$, and $C = 0.66$.

(An animation of this figure is available in the online article.)

Note that these results here are specific to the chosen planet–star system. Any change in the system's geometry or parameters requires the grid to be regenerated.

We then introduce spot contrast, $C$, which can vary from 0 to 1, expanding our grid into 4D parameter space. For each spatial configuration, we add a range of contrasts, multiplying the total grid size by the number of contrast values. The instantaneous change in-transit depth due to the spot crossing, $\Delta D(t) = -g(t)C\epsilon D_{\rm true}$, from Equation (9) allows us to then define the spot-bump amplitude, $\Delta D_{\rm spot}$, as half the height (FWHM) of the SCE, for every spot configuration in our grid

$$\Delta D_{\rm spot} = \frac{1}{2}\max|\Delta D(t)| = \frac{1}{2} g_{\rm max} C \epsilon D_{\rm true} \tag{16}$$

where $g_{\rm max}$ is the peak fraction of the planet's shadow covered by spots.

#### 6.2.2. Using Lightcurve Observables to Narrow the Degeneracy Grid

Given an observed SCE, we can extract the measured quantities, $t_{\rm spot}$, $\Delta t_{\rm spot}$, and $\Delta D_{\rm spot}$—each with associated observational uncertainties. We then search the degeneracy grid to identify only those spot crossings whose predicted observables lie within those error bounds. This filtering step yields a subgrid of degenerate solutions, each with a spot contrast, radius, latitude, and longitude consistent (within some uncertainty) with the observed event. The effect of varying the spot-crossing observables on the degeneracy space is demonstrated in the animation in Figure 12.

### 6.3. Assumptions

We made several significant assumptions in this section. Unlike the Gaussian-smoothed stellar surface from `starry` utilized, in Section 3 we simplify our model to a circular spot with hard edges. In reality, starspots are unlikely to be perfectly circular with uniform contrast, but this simplification is the basis for several other commonly used tools, including `spotrod` (B. Béky et al. 2014), `fleck` (B. Morris 2020), and `STSP` (B. M. Morris et al. 2017). By employing a geometric approach rather than the full spherical harmonic representation (as in `starry`), we reduce computation time and increase the spatial resolution we can model. The hard-edged circular spot model corresponds most closely to a `starry` implementation with zero smoothing. In this limit, `starry` lightcurves exhibit ringing artifacts due to sharp brightness discontinuities. As a result, small but non-negligible discrepancies arise between the predicted relationships for SCE observables and underlying spot parameters when comparing our geometric model to `starry`-based results.

When generating the degeneracy grids, we assume that $t_{\rm spot}$ is the midpoint of each spot crossing. When circular spots are projected into ellipses and moved away from the stellar equator they are foreshortened along the line of sight, which can skew the ingress and egress of the spot-crossing signal and lead to asymmetric SCEs when both $\lambda_{\rm spot} \neq 0$ and $\phi_{\rm spot} \neq 0$. This effect is most prominent for spots near the stellar limbs. We allow this asymmetry within our framework ($g(t_{\rm spot}) \neq g_{\rm max}$ necessarily), though this effect is not present in Figure 12 because the planet is always crossing the spot centrally and at the equator. In reality, this asymmetry may be compounded by unknown spot complexities (e.g., a noncircular shape, nonuniform contrast, or locally irregular stellar surface brightness such as a strong limb-darkening gradient across the spot).

An SCE will provide information about a singular spot, or spot complex, at that point in time. Modeling this event in isolation does not provide information about spot evolution, rotation, or the presence of other surface heterogeneities. However, a single spot will provide a lower limit for the spot-coverage fraction and, if we assume all spots on a stellar surface share the same temperature, a global spot contrast. Both of these parameters are necessary for then constraining stellar contamination in transmission spectroscopy.

## 7. Kepler-51 d: A Starspot-crossing Event Observed by JWST

We apply this degeneracy-mapping technique to the real JWST lightcurve of Kepler-51d from Program GO-2571 (PI Libby-Roberts). The GO-2571 program provides a useful test case because it involves only a single high-precision transit with a clear SCE; therefore, any insights into the stellar surface are limited by how well we can model this isolated event.

Kepler-51 is a young Sun-like ($M = 0.96\,M_\odot$, $R = 0.87\,R_\odot$) star that hosts three extremely low density, "super-puff" transiting planets, including the Saturn-sized Kepler-51 d ($R = 9.32 \pm 0.18\,R_\oplus$), as well as a smaller nontransiting planet (J. H. Steffen et al. 2013; K. Masuda 2014; K. Masuda et al. 2024). A single transit of Kepler-51 d was obtained with JWST NIRSpec-PRISM on 2023 June 26 UT (J. E. Libby-Roberts et al. 2026).[7] This data was reduced using multiple pipelines, including Eureka! (T. J. Bell et al. 2022), revealing at least one SCE (shown in Figure 13(a)). There is a clear event midtransit, as well as a potential smaller event during ingress.

As described in J. E. Libby-Roberts et al. (2026), the white lightcurve for a transit and SCE was fitted by combining a second-order polynomial with two different spot modeling tools and sampling methods: (*a*) `starry` and MCMC (within `chromatic_fitting` ) and (*b*) `spotrod` and Dynesty.

### 7.1. Transit Depth Uncertainty Inflation

Kepler-51 d's reduced white lightcurve has median per-point uncertainty $\sigma = 350$ ppm at the native cadence of 2.9 s (8.5 hr in-transit, 5 hr out-of-transit). Method (*b*) fits the unbinned white lightcurve. Conversely, for the `starry` (*a*) method, the white lightcurve was binned before fitting to a cadence of 2 minutes yielding average per-point uncertainties, $\sigma$, of ~61 ppm, 255 data points in-transit, and 193 data points out of transit. If we assume transit depth uncertainties scale as $\sigma\sqrt{1/N_{\rm in} + 1/N_{\rm out}}$, we would expect a theoretical transit depth uncertainty of ~6 ppm, and, from Figure 10, an MCMC-recovered depth uncertainty of ~9 ppm. From sampling, the transit depth is recovered as (a) $9673 \pm 19$ ppm and (b) $9374 \pm 99$ ppm. Therefore, we find that the theoretical transit depth uncertainties inflate by factors of (a) 3.4 and (b) 17.0 for $\log(\sigma/\sqrt{N}) \sim -5.4$ across the two scenarios. From Figure 10, we might have predicted ~10–30x inflation. There are a few potential reasons for overestimating Kepler-51d's depth uncertainty inflation. The recovered spot on Kepler-51 is low contrast (spot contrasts recovered from 0.087 to 0.096), which should have a low impact on depth recovery. Additionally, as mentioned in Section 5.1, our injection-recovery tests were for a broad range of SCEs, not specifically to Kepler-51d.

### 7.2. Degeneracy Space

We also applied our method for exploring degeneracies, described in Section 6, to the larger SCE in Kepler-51d's lightcurve. We generated a degeneracy grid for Kepler-51d using the following parameters that J. E. Libby-Roberts et al. (2026) used in, and derived from, the `starry` fit: stellar mass, $M_* = 0.985\,M_\odot$, stellar radius, $R_* = 0.862\,R_\odot$, stellar rotation period, $P_{\rm rot} = 8.222$ days, orbital period, $P = 130.185$ days, eccentricity, $e = 0$, orbital inclination, $i = 89\overset{\circ}{.}88$, epoch $t_0 = 2460121.85$, and quadratic limb-darkening coefficients [0.20, 0.39]. We created our degeneracy grid limiting our spot radii from 1° to 30°, latitudes from −70° to +30° (due to the nonzero impact parameter), and longitudes from −100° to +100°, using the same resolution as before of 1°. This results in a grid of 609,030 points, of which 241,424 feature an SCE.

Using the residuals between the lightcurve and the transit model (without the SCE) in Figure 13(b), we determined $\Delta t_{\rm spot}$, $t_{\rm spot}$, and $\Delta D_{\rm spot}$ by fitting a Gaussian model with least squares, extracting its mean $\mu$, standard deviation $\sigma$ and height $A$ with respective uncertainties. By fitting a Gaussian model, we make the reasonable assumption that the SCE is symmetric since the spot is not near the stellar limbs (see Section 6.3). From this Gaussian, we take $\Delta D_{\rm spot} = A/2$, $t_{\rm spot} = \mu$, and $\Delta t_{\rm spot} = {\rm FWHM} = 2\sigma\sqrt{2\ln 2}$. For each quantity we propagate the uncertainties from least squares and use a very conservative $10\sigma$ for our observable uncertainties. Therefore, for Kepler-51d's larger SCE, we have $t_{\rm spot} = 2460121.871 \pm 0.008$ day, $\Delta t_{\rm spot} = 0.06 \pm 0.02$ day, and $\Delta D_{\rm spot} = 433 \pm 114$ ppm. In Figures 13(a) and (b), we show the white lightcurve with the three SCE observables marked in purple and the Gaussian model in yellow.

We now use the three SCE observables to constrain the degeneracy space. The measured SCE epoch, $t_{\rm spot}$, limits the spot longitude to $7 \leqslant \lambda_{\rm spot} \leqslant 15°$ (Figure 13(c)), while the SCE duration, $\Delta t_{\rm spot}$, requires a spot radius of at least $R_{\rm spot} \geqslant 9°$ (Figure 13(d)). Constraining the spot latitude is more complex due to the strong degeneracy between latitude and radius (Figure 13(e)). The measured SCE bump size, $\Delta D_{\rm spot}$, provides a lower limit on the spot contrast of $\geqslant 0.06$ (Figure 13(f)).

In Figures 13(c)–(f), we overplot the spot parameters derived from the `starry` (orange, $R_{\rm spot} = 12\overset{\circ}{.}0$, $\lambda_{\rm spot} = 7\overset{\circ}{.}7$, $\phi_{\rm spot} = -9\overset{\circ}{.}7$, $C = 0.087$) and `spotrod` (cyan, $R_{\rm spot} = 17\overset{\circ}{.}4$, $\lambda_{\rm spot} = 8\overset{\circ}{.}7$, $\phi_{\rm spot} = -30\overset{\circ}{.}3$, $C = 0.089$) spot fits in J. E. Libby-Roberts et al. (2026). To understand how the various spot-modeling frameworks are sampling the degeneracy space, we plot the respective posterior traces (J. E. Libby-Roberts 2025, private communication) in Figure 13(e).

To demonstrate the degeneracy modeling, we also fit this lightcurve with `fleck` twice: once with uninformative priors and once implementing priors from carving out the degeneracy space with geometry ($7 \leqslant \lambda_{\rm spot} \leqslant 15°$, $R_{\rm spot} \geqslant 9°$ and $C \geqslant 0.06$). While we do not expect the priors to significantly change the MCMC sampling because, even with uninformative priors, it will eliminate unphysical solutions during its run, restricting the priors may prevent the samples from inefficiently wandering across vast degenerate spaces. Though we see similar posterior regions for both fits in Figure 13(e), there are two marked improvements when adding informed priors. First, we reach our MCMC convergence criteria (50× autocorrelation time) faster, within 6000 steps compared to 10,000 steps for uninformative priors. Second, our posterior appears to sample a larger proportion of the degeneracy space, now overlapping with the spotrod samples, perhaps due to more efficient sampling. We include the spot parameters from both fleck fits in Figure 13 with uninformative priors in light green ($R_{\rm spot} = 25^\circ\!.1$, $\lambda_{\rm spot} = 7^\circ\!.4$, $\phi_{\rm spot} = 5^\circ\!.4$, $C = 0.159$) and geometric-bounded priors in dark green ($R_{\rm spot} = 16^\circ\!.6$, $\lambda_{\rm spot} = 7^\circ\!.8$, $\phi_{\rm spot} = -5^\circ\!.3$, $C = 0.09$).

[7] The JWST/NIRSpec-PRISM data of Kepler-51d may be obtained from the MAST archive at DOI: 10.17909/s4bw-fh78.

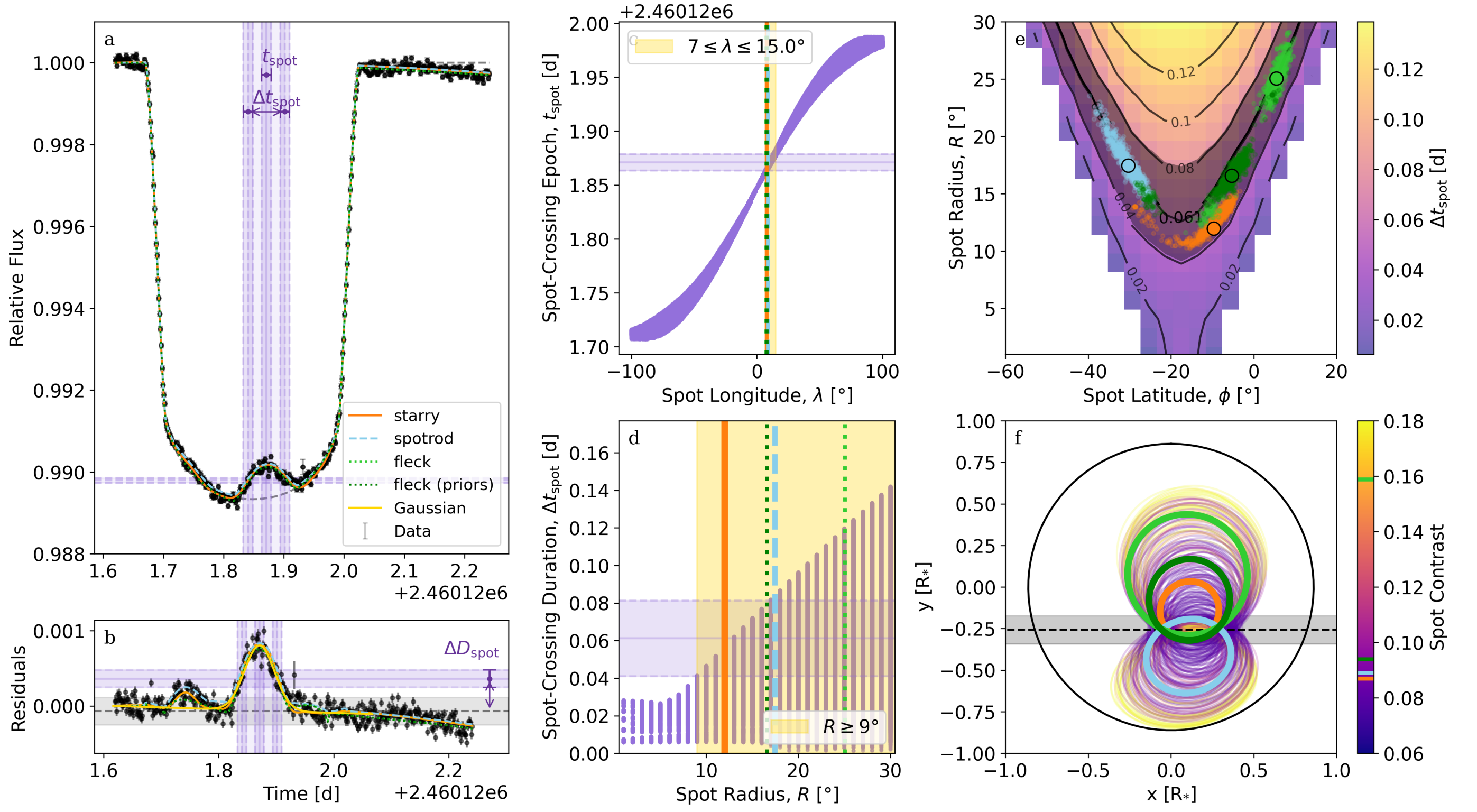

**Figure 13.** (a) JWST white lightcurve for Kepler-51d. Best-fit transit and spot models (starry in orange, spotrod in blue, fleck with uninformative priors in light green, and fleck with geometric-based priors in dark green), with a second-order polynomial. The dotted black line is the same transit model without the spot crossings, rescaled to the same contaminated transit depth as the starry model. We label $t_{\rm spot}$, the spot-crossing epoch, and $\Delta t_{\rm spot}$, the spot-crossing duration, along with their uncertainties. (b) Residuals between white lightcurve and scaled transit model. We label the spot-crossing bump size, $\Delta D_{\rm spot}$.We include the Gaussian model used to extract the spot-crossing observables in yellow. (c) $t_{\rm spot}$ as a function of spot longitude, $\lambda$, for the degeneracy grid is plotted in purple. The observed $t_{\rm spot}$ and uncertainty are highlighted by the horizontal line. $\lambda$s for the grid spots that match the $t_{\rm spot}$ are shown by the vertical shaded yellow region, corresponding to $7° \leqslant \lambda \leqslant 15°$. We include the best-fit starry (orange) and spotrod (blue) values from J. E. Libby-Roberts et al. (2026) and fleck (green) from this work. (d) Similarly to (c), we plot $\Delta t_{\rm spot}$ as a function of the spot radius, $R$. The grid spots that match the observed $\Delta t_{\rm spot}$ have radii $\geqslant 9°$. (e) Heatmap showing the degenerate relationship between $R$ and spot latitude, $\phi$, for the same $\Delta t_{\rm spot}$ (contours and colorbar). The darkest contour is the observed $\Delta t_{\rm spot}$ and the shaded gray region its uncertainty. The MCMC samples for starry and fleck and Dynesty samples for spotrod are overplotted, each exploring a subset of the degenerate region. (f) The resulting grid spots that match the observed $t_{\rm spot}$, $\Delta t_{\rm spot}$, and $b_{\rm spot}$, plotted on top of Kepler-51. The transit chord is shown by the black dotted line and shaded gray region. The colorbar indicates the spot contrast, $C$. We mark the best-fit spots for each fitting routine on the star and their contrasts on the colorbar.

The spotrod and fleck-derived spot properties all fall easily within bounds derived from the observables alone, whereas the starry latitude–radius relationship (Figure 13(e)) only matches these bounds with conservative $10\sigma$ uncertainties on the observables. If we were to reduce these, for example, to $5\sigma$, a fraction of the starry posterior samples would be excluded. This discrepancy is not surprising due to the underlying model differences. Our degeneracy exploration assumes a hard-edged circular spot—matching the assumptions in spotrod and fleck—while starry smooths the stellar surface, effectively "spreading" the spot and extending the spot-crossing duration. This allows us to retrieve spots smaller than we might expect from the same observables. This likely explains why the starry posterior trend in Figure 13(e) generally follows a similar shape to the spotrod and fleck samples but a offset by a couple of degrees.

## 8. Discussion

In this work, we aim to set up general frameworks to better quantify the impact of SCEs on transit lightcurves. First, we inject and recover SCEs in a simple transit lightcurve (Sections 3 and 4), performing full MCMC sampling for five scenarios (Section 5).

In 80% of our injection and recoveries of synthetic spot crossings, we recover the true transit depth ($D = 10{,}000$ ppm) to within 0.6% or 253 ppm, and the spot contrasts, radii, longitudes, and latitudes to within $0^\circ\!.15$, $5^\circ\!.0$, $1^\circ\!.0$, and $8^\circ\!.7$, respectively. We expected to recover the longitude ($\sim x$-position) well given that it corresponds directly to the SCE epoch. The main exception is when we have a spot on the stellar limb. Due to projection effects, the range of longitudes compatible with a spot crossing during ingress or egress is larger than in the center of the transit, and detections of spot crossings on the limb are difficult to disentangle from limb-darkening, occasionally resulting in fitting an unocculted spot at a random location. Due to uncertainty in the lightcurve and strong degeneracies, the latitude of the spots ($\sim y$-position) is very poorly constrained. Since there is degeneracy between latitude, radius, and contrast, we recover the radii and contrasts less precisely than the longitudes or transit depths.

We find that if spots cause contaminations $\epsilon > 1.3\%$, fitting for a spot crossing in a single-wavelength lightcurve improves the recovery of the transit depth in over 95% of cases compared to masking and avoids the significant uncertainties introduced by discrepant and inaccurate stellar spectral models. B. V. Rackham & J. de Wit (2024) found that differences between stellar model grids for M-dwarfs (on average ∼200 ppt) dominated the noise budget of their planets' transmission spectra. This uncertainty is larger than all of the recoveries (including an order of magnitude larger than most recoveries) from fitting high SNR spot crossings in Section 4.2. B. V. Rackham & J. de Wit (2024) found that the disagreement between stellar models varied as a function of wavelength, with larger discrepancies toward the visible; however, we do not explore the effect of wavelength in this work. Additionally, if we can extract the spot contrast accurately from single lightcurves by modeling SCEs, we can extend this to then deriving a spot contrast spectrum. An empirical contrast spectrum could then provide an alternate method of fitting for the TLSE in transmission spectra directly, without having to rely on stellar models.

This work adds to growing evidence for performing occulted spot analysis to characterize the contamination in our atmospheric inferences (e.g., M. Fournier-Tondreau et al. 2024; J. E. Libby-Roberts et al. 2026). However, within our retrieval framework, we note that the parameters for low SNR (<4) spot crossings, and spots on stellar limbs, were not well recovered. Additionally, the transit depth for very small or low contrast spots was often overcorrected due to the significant degeneracies. Therefore, we recommend caution when modeling spot crossings in the cases of small, low SNR SCEs, though fortunately the influence of these spots on the final transmission spectrum is likely minimal.

Within our framework we find that SCEs, even when fitted, inflate the uncertainties on recovered transit depths. For JWST-like precisions, this could inflate transmission spectra uncertainties on the order of 10–100x. Therefore, in the case of known "spotty" stars we recommend a conservative approach of simulating the impact of occulted spots when proposing for telescope time to ensure science goals will be achieved.

Next, we introduced an approach for exploring the degeneracy space for SCEs directly using lightcurve observables (spot-crossing epoch, duration, and bump size) in Section 6. Using degenerate spot solutions consistent with these observables allows us to place variably constraining bounds on spot contrast, size, and location. We apply this method to the spot crossing in the JWST lightcurve of Kepler-51d. We extract limits on spot properties from the observables and compare to the results from `spotrod` and `starry` presented in J. E. Libby-Roberts et al. (2026). We perform additional fits to Kepler-51 d's lightcurve with `fleck` to compare the three commonly used tools. We fit with `fleck` twice: first with uninformative priors and then with the spot parameter bounds derived from SCE observables, which we find to converge faster and more efficiently explore the posterior space.

We find that for a similar hard-edged circular model to the one implemented in our work (e.g., for `spotrod` and `fleck`), these results are consistent, but we find slight discrepancies with `starry`'s smoothed spot model. The reliance of spot parameters on the assumed morphology is unsurprising but concerning. The Sun's spots have two distinct regions: a dark umbra core and surrounding penumbra region; however, relatively little is known about spot structure on other stars, especially M-dwarfs. When we observe an SCE we do not know local gradients of photospheric temperature or the relative proportions of umbral and penumbral regions a planet is transiting. Therefore, our recommendation for the community is to consider multiple tools with different spot morphologies (e.g., hard-edged circles, harmonics, or pixels) when deriving starspot properties and discuss the consistencies and discrepancies between them.

While fitting the SCE directly will provide better spot constraints than our simplified three-observable approach, all of the spot-fitting tools used for Kepler-51 seem to sample different regions of the degeneracy space. All methods derive similar longitudes and three find consistent contrasts, but all significantly diverge in latitude and size. These differences may stem in part from model assumptions, but could also reflect limitations in the sampling, with each method potentially getting trapped in different degenerate solutions. Systematically exploring the degeneracy space offers a valuable framework for interpreting such discrepancies and for understanding the limitations inherent to different spot-fitting tools. Finally, assuming that consistent models are used, geometric priors can help to improve sampling efficiency.

## 9. Conclusion

Stellar contamination—through occulted and unocculted active features—remains a major obstacle to accurately characterizing the atmospheres of planets orbiting cool stars. To date, detailed spectrophotometric studies of such hosts (e.g., Z. K. Berta et al. 2011; B. Rackham et al. 2017; E. Ducrot et al. 2018; J. D. Lothringer et al. 2018; Z. Zhang et al. 2018; O. Lim et al. 2023; E. M. May et al. 2023; K. A. Bennett et al. 2025) have been relatively limited, yet understanding the stellar surface is crucial for disentangling stellar and planetary signals. The upcoming small satellite mission Pandora (E. V. Quintana et al. 2024), set to launch in 2025, will combine photometry simultaneously with near-infrared spectroscopy to decouple stellar and planetary signals for 39 low-mass exoplanet hosts and provide new understandings of their relationship. This dataset is expected to significantly advance our understanding of stellar contamination and its impact on transmission spectra.

Direct transit scans of active regions, such as starspots, provide a unique window into the stellar surface. Though spot parameters are inherently degenerate, here we demonstrate how observations of SCEs can successfully constrain spot properties in most cases. Our results indicate that fitting for SCEs is almost always preferable to masking them, and that spot contrast can be robustly inferred without relying on spectral stellar models. Furthermore, by inferring the spot contrast from occulted features empirically we can correct the transmission spectra for additional unocculted features of the same temperature by fitting for only the remaining coverage fraction, rather than fitting for both coverage fraction and spot contrast temperature (again relying on stellar models). Therefore, SCEs offer a pathway for correcting wavelength-dependent contamination in transmission spectra directly from the data. Although our analysis focused on single-wavelength transits, this approach and the results can be easily extended to any wavelength, laying groundwork for more comprehensive stellar contamination exploration in future wavelength-resolved studies. The wealth of high-precision time-resolved spectra now arriving from JWST makes it possible not just to mitigate stellar noise but exploit it to reveal insights into stellar activity and the impact on planetary atmospheres.

## Acknowledgments

We thank the reviewer for their invaluable feedback on this manuscript. We thank Jessica Libby-Roberts for many insightful conversations on Kepler-51d, and for sharing their data and results. This material is based upon work supported by the National Science Foundation under grant No. 1945633. This work is based in part on observations made with the NASA/ESA/CSA James Webb Space Telescope. The data were obtained from the Mikulski Archive for Space Telescopes at the Space Telescope Science Institute, which is operated by the Association of Universities for Research in Astronomy, Inc., under NASA contract NAS 5-03127 for JWST. These observations are associated with program JWST-GO-2571 (PI Libby-Roberts).

*Facility:* JWST (NIRSpec).

*Software:* `chromatic` (Z. Berta-Thompson et al. 2025), `chromatic_fitting` (C. Murray et al. 2025), `starry` (R. Luger et al. 2019), `shapely` (S. Gillies et al. 2024), `PyMC3` (J. Salvatier et al. 2016), `astropy` (Astropy Collaboration et al. 2013 2018, 2022), `numpy` (C. R. Harris et al. 2020), `matplotlib` (J. D. Hunter 2007).

## Appendix
## Injection-recovery Outliers

Here we present two edge cases of our injection-recovery tests. First, in Figure A1, we show the five largest outliers in recovered depth. All show spot contrasts and radii at the edges of the allowed parameter space, with spots mostly covering the stellar limb. Second, in Figure A2, we present the five tests with the most over-corrected recovered transit depths. All cases have relatively low spot contrasts and the injected spot radius for three approaches our lower limit of 5°. Despite high SNRs, the injected and recovered lightcurves are visually identical, highlighting how parameter degeneracies can bias spot recovery, even with excellent data precision.

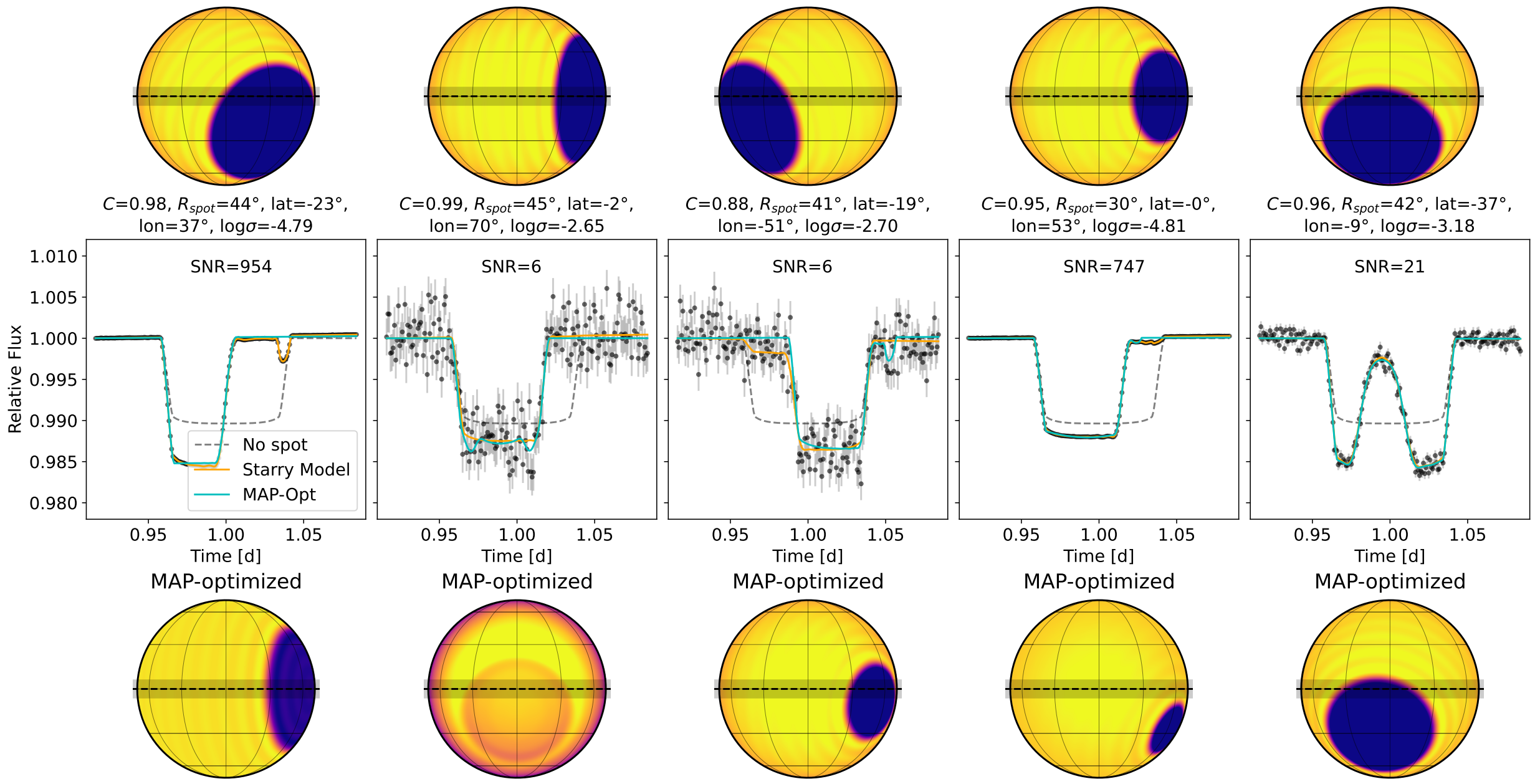

**Figure A1.** The five largest outliers in recovered depth (in descending order). The layout of this plot is the same as Figure 1. In each case we see unphysically large spot contrasts and sizes, and a poor recovery of the true spot, especially for spots on the stellar limb.

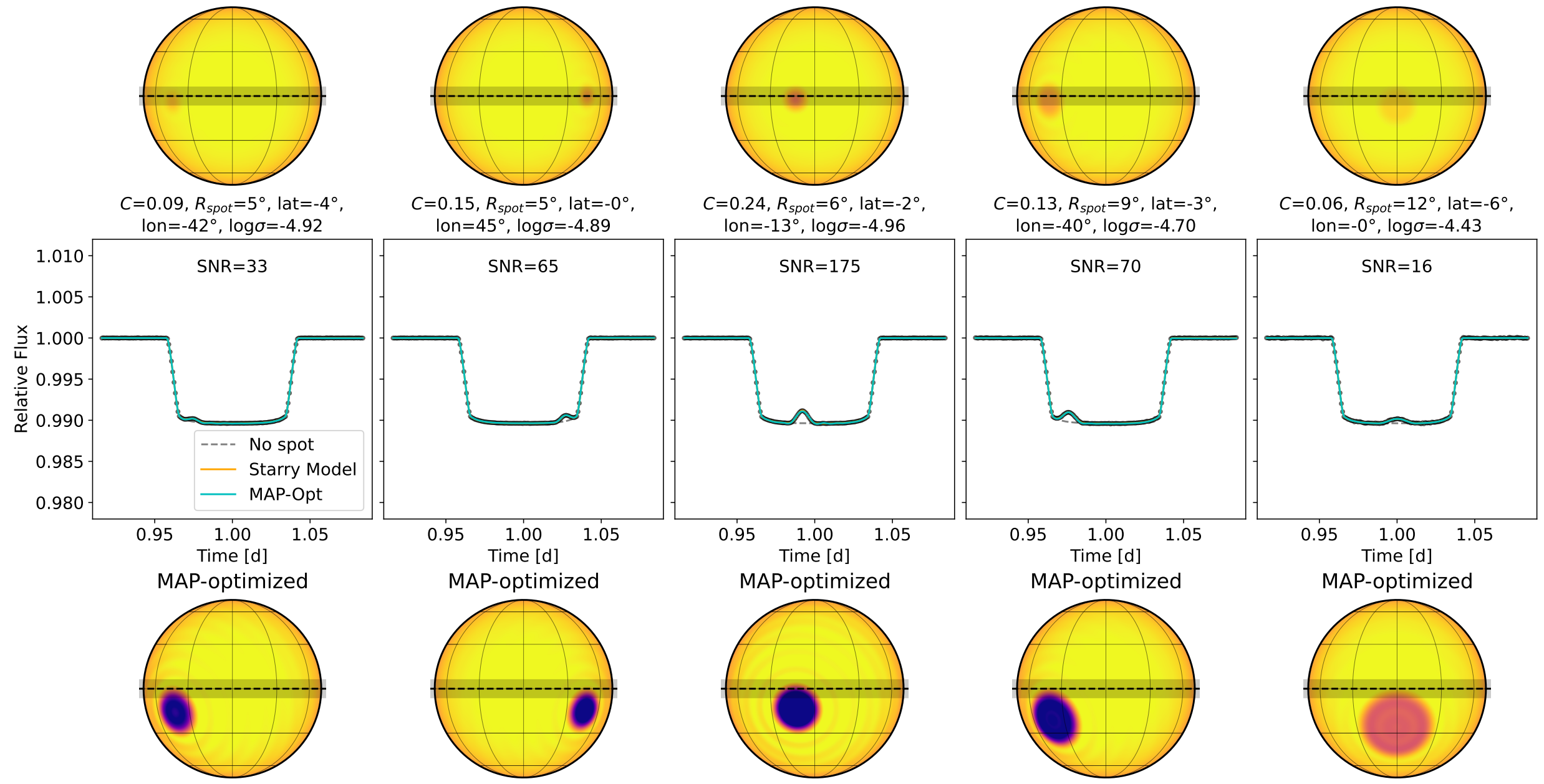

**Figure A2.** The five scenarios with the smallest recovered (most overcorrected) transit depths. The layout of this plot is the same as Figure 1. In each case a degenerate solution, that produces an indistinguishable lightcurve model, is preferred with a larger size and contrast than the original

## ORCID iDs

Catriona A. Murray https://orcid.org/0000-0001-8504-5862
Zachory Berta-Thompson https://orcid.org/0000-0002-3321-4924